\newcommand{\crossed}{>\!\!\!\triangleleft}

\newcommand{\dbcocrossed}{\blacktriangleright \!\! \blacktriangleleft}
\newcommand{\brcrossed}{>\!\!\!\triangleleft \!\!\! \cdot}

\newcommand{\trr}{\triangleright}
\newcommand{\U}{{U} ({\mathfrak su}(2))}
\newcommand{\RL}{{\mathbb R}^3_\lambda}
\newcommand{\UL}{{\mathbb R}^3_\lambda}
\newcommand{\CSU}{{\mathbb C} (SU(2))}

\newcommand{\DU}{D(U ({\mathfrak su}(2)))}

\newcommand{\UQ}{U_{q} ({\mathfrak su}(2))}
\newcommand{\BSU}{BSU_q (2)}
\newcommand{\su}{{\mathfrak su}(2)}
\newcommand{\dsl}{\partial\kern-7pt /}

\newcommand{\h}{{\scriptstyle\frac{1}{2}}}
\newcommand{\extd}{{\rm d}}
\newcommand{\del}{\partial}

\newcommand{\isom}{{\cong}}
\newcommand{\eps}{{\epsilon}}
\newcommand{\tens}{\mathop{\otimes}}
\newcommand{\la}{{\triangleright}}

\newcommand{\Ad}{{\rm Ad}}
\newcommand{\id}{{\rm id}}
\newcommand{\<}{\langle}
\renewcommand{\>}{\rangle}
\newcommand{\End}{{\rm End}}

\newcommand{\R}{\mathbb{R}}
\newcommand{\C}{\mathbb{C}}
\newcommand{\Z}{\mathbb{Z}}
\newcommand{\CR}{{\mathcal R}}
\newcommand{\CI}{{\mathcal I}}
\newcommand{\CO}{{\mathcal O}}
\newcommand{\CE}{{\mathcal E}}
\renewcommand{\u}{\mbox{\bf u}}
\renewcommand{\L}{\mbox{\bf l}}
\newcommand{\M}{\mbox{\bf m}}
\newcommand{\N}{\mbox{\bf n}}
\renewcommand{\l}{\lambda}
\newcommand{\dd}{\mbox{\bf d}}

\newcommand{\bray}{\begin{eqnarray}}
\newcommand{\eray}{\end{eqnarray}}
\newcommand{\ea}{\end{array}}
\newcommand{\proof}{{\bf Proof\ }}
\newcommand{\eproof}{$\quad \diamond$\bigskip}

\newcommand{\eqn}[2]{\begin{equation}#2\label{#1}\end{equation}}

\documentclass[11pt]{article}
\usepackage{amssymb,amsmath,epsfig}
\newtheorem{lemma}{Lemma}[section]

\newtheorem{theorem}[lemma]{Theorem}

\begin{document}\baselineskip 18pt

{\ }\qquad \hskip 4.3in \vspace{.2in}

\begin{center} {\LARGE NONCOMMUTATIVE GEOMETRY OF ANGULAR MOMENTUM
SPACE $U(\su)$} \\ \baselineskip 13pt{\ }\\
{\ }\\
Eliezer Batista\footnote{Supported by CAPES, proc. BEX0259/01-2.}
\\{\ }\\
Dep. de Matem\'atica, Universidade Federal de Santa Catarina\\
CEP: 00.040-900, Florianopolis-SC, Brazil\\{\ }\\ +\\{\ }\\Shahn
Majid\footnote{Royal Society University
Research Fellow}\\ {\ }\\
School of
Mathematical Sciences\\ Queen Mary, University of London\\
Mile End Rd, London E1 4NS, UK
\end{center}
\begin{center}
April 2002 -- revised May 2002
\end{center}
%\vspace{10pt}

\begin{quote}\baselineskip 13pt
\noindent{\bf Abstract} We study the standard angular momentum
algebra $[x_i,x_j]=\imath\lambda \eps_{ijk}x_k$ as a
noncommutative manifold $\RL$. We show that there is a natural 4D
differential calculus and obtain its cohomology and Hodge
* operator. We solve the spin 0 wave equation and some aspects of
the Maxwell or electromagnetic theory including solutions for
a uniform electric current density, and we find a natural
Dirac operator $\dsl$. We embed $\RL$ inside a 4D noncommutative
spacetime which is the limit $q\to 1$ of q-Minkowski space and
show that $\RL$ has a natural quantum isometry group given by the
quantum double $\CSU \crossed U(su_2)$ which is a singular limit
of the $q$-Lorentz group. We view $\R_\lambda^3$ as a collection
of all fuzzy spheres taken together. We also analyse the
semiclassical limit via minimum uncertainty states
$|j,\theta,\phi\>$ approximating classical positions in polar
coordinates.
\end{quote}

\baselineskip 18pt

\section{Introduction}

There has been much interest in recent years in the possibility
that classical space or spacetime itself (not only phase space) is
in fact noncommutative and not a classical manifold. One simple
model where \eqn{kappamink}{[x_i,t]=\imath\lambda x_i} has already
been shown \cite{AmeMa:wav} to have physically measurable effects
even if $\lambda\sim 10^{-44}$ seconds (the Planck time). So such
a conjecture is not out of reach of experiment {\em even if} the
noncommutativity is due to quantum gravity effects. Such
noncommutativity of spacetime, if verified, would amount to a new
physical effect which could be called `cogravity' because it
corresponds under non Abelian Fourier transform to curvature in
momentum space\cite{Ma:jmp}. We are usually familiar with this
correspondence the other way around, i.e. on a curved space such
as a sphere the canonical momenta (angular momentum) form a
noncommutative algebra \eqn{J}{ [J_a,J_b]=\imath
\eps_{abc}J_c.,\quad a,b,c=1,2,3} where $\eps_{abc}$ denotes the
totally antisymmetric tensor; if one believes in Born reciprocity
then one should also allow the theoretical possibility of a sphere
in momentum space, which would corresponds to the algebra \eqn{x}{
[x_a,x_b]=\imath\lambda\eps_{abc}x_c.} This is the algebra $\RL$
which we study in this paper from the point of view of the $x_i$
as coordinates of a noncommutative position space. We insert here
a parameter $\lambda$ of length dimension. The physical relevance
of this algebra hardly needs to be justified, but we note some specific
applications in string theory and quantum gravity in the discussion below.
 There are also possible
other contexts where a noncommutative spacetime might be a good
effective model, not necessarily connected with gravity and indeed
this is an entirely independent (dual) effect. 

Also from a mathematical point of view, the algebra (\ref{x}) is a
standard example of a formal deformation quantisation, namely of
the Kirillov-Kostant Poisson bracket on ${\mathfrak su}_2^*$ in
the coadjoint
orbit method \cite{kirillov}. We identify $\su^*$ as the vector
space  $\R^3$ with basis $J_a^*$, say, dual to the $J_a$. Then
among the algebra of suitable (polynomial) functions $\C(\R^3)$ on
it we identify the $J_a$ themselves with the `coordinate
functions' $J_a(v)=v_a$ for any $v\in \su^*$ with component $v_a$
in the $J_a^*$ direction. These generate the whole coordinate
algebra and their Poisson bracket is defined by \[
\{J_a,J_b\}(v)=v([J_a,J_b]),\quad \forall v\in \su^*.\] Hence when
viewed as functions on $\R^3$ the Lie algebra generators have a
Poisson bracket given by the Lie bracket. Their standard
`quantisation' is evidently provided by (\ref{x}) with deformation
parameter $\lambda$.

Our goal in the present work is to use modern quantum group
methods to take this further by developing the noncommutative
differential geometry of this quantum space at the level of scaler
fields, forms and spinors, i.e. classical field theory. We will
solve wave equations etc. and generally show that physics is fully
possible on $\RL$. Note that the earlier example (\ref{kappamink})
above was also of `dual Lie' type but there the Lie algebra was
solvable whereas the $\su$ case that we address here is at the
other extreme and very much harder to work with. We expect our
methods to extend also to $U(g)$ for other simple $g$.

The paper begins in Section 2 with some mathematical preliminaries
on quantum group methods and noncommutative geometry. As a quantum
group, $\RL\isom U(\su)$ (the enveloping Hopf algebra) which means
that at the end of the day all computations can be reduced to the
level of $\su$ and Pauli matrices. One of the first things implied
by quantum group theory is that $\RL$ has an isometry quantum
group given by the Drinfeld quantum double $D(U(\su))$ and we
describe this first, in Section 3. A suitable Casimir of this
induces a scalar wave operator $\square$ and we also
describe spherical harmonics $Y_l^m$ dictated by action of
rotations. This theory could be called the `level 0'
noncommutative geometry where we think of the space through its
symmetries rather than it differential structure.

In Section~4 we start the noncommutative differential geometry,
introducing a natural differential calculus on $\RL$. The
cotangent directions or basic forms are given literally by Pauli
matrices plus an additional generator $\theta$: \eqn{dxa} {\extd
x_a={1\over 2} \sigma_a,\quad \theta= \sigma_0} where
$\sigma_0=\id$ (the identity matrix). There are also
non-commutation relations between functions and 1-forms:
\eqn{dxacomm}{ x_a \extd x_b = (\extd x_b) x_a +\frac{\imath\l}{2}
\epsilon_{abc} \extd x_c +\frac{\l}{4} \delta_{ab} \theta,\quad
x_a \theta = \theta x_a + \l \extd x_a.} Some other calculi are
mentioned in Appendix A for comparison but in fact this 4
dimensional one appears to be the most reasonable one. The extra
$\theta$ direction turns out to generate the cohomology i.e. is
not $\extd$ of anything in $\RL$. We interpret it as a local time
direction in the same spirit as in a different model
\cite{xavier1}.

In Section~5 we introduce a  Hodge * operator and solve the
resulting wave equations for spin 0 and spin 1 (the Maxwell
equations). We also find a natural Dirac operator for spin $1/2$.
Among the solutions of interest are plane waves obeying
\[ \square
e^{\imath k\cdot x} =-\frac{1}{\l^2}\left\{ 4 \sin^2 \left(
\frac{\l \mid k\mid}{2} \right) +\left( \cos \left( \frac{\l \mid
k\mid}{2} \right) -1\right)^2 \right\} e^{\imath k\cdot x} .
\]
for momentum $k\in \R^3$. Among spin 1 solutions is a uniform
electric current density in some direction and magnetic field increasing
with normal distance. This is computationally the easiest case; we 
expect that the theory should similarly allow more
conventional decaying solutions. In Section~6 we briefly consider
quantum spheres $S^2_\lambda$ inside $\R^3_\lambda$ by setting
$\sum_i x_i^2=1$. These are then the usual quantization of
coadjoint orbits in $\su^*$ (as opposed to all of $\su^*$ as
described above) and we show that they inherit a 3-dimensional
differential geometry. This case could be viewed as a slightly
different approach to fuzzy
spheres\cite{madore,bal,HLS,pinzul,ramgoolan} that is more adapted to
their classical limit $\lambda\to 0$. Fuzzy spheres also arise as
world volume algebras in string theory\cite{ARS} hence it would be
interesting to develop this point of contact further. In our case
we obtain a 3D differential calculus on $S^2_\lambda$.

In Section~7 we explain the origin of the $\theta$ direction as
the remnant of the time direction $\extd t$ of a standard
4-dimensional noncommutative spacetime $\R_q^{1,3}$ in a certain
scaling limit as $q\to 1$. In the $q\ne 1$ setting the theory is
much more nonsingular and there is a full $q$-Lorentz symmetry
already covered in the $q$-deformation
literature\cite{CWSS,majidlorentz,majid}. On the other
hand, as $q\to 1$ we obtain either usual commutative Minkowski
space or\cite{majidlorentz} we can make a scaling limit and obtain
the algebra \eqn{homogsu}{ [x_a,x_b]=\imath c t\,
\eps_{abc}x_c,\quad [x_a,t]=0} where the parameter $c$ has
dimensions of velocity. Mathematically this is homogenised
$\widetilde{U(\su)}$ and we see that is projects onto our above
algebra (\ref{x}) by sending $ct\to \lambda$. This algebra
(\ref{homogsu}) is not itself a good noncommutative Minkowski
space because the $q$-Lorentz group action becomes singular as
$q\to 1$ and degenerates into an action of the above quantum
double isometry group. On the other hand it is the boundary point
$q=1$ of a good and well-studied noncommutative Minkowski space.

The paper concludes in Section~8 with a proposal for the
interpretation which is needed before the noncommutative geometry
can be compared with experiment. In addition to a normal ordering
postulate (i.e. noncommutative $f(x)$ are compared with classical
ones only when normal ordered) along the lines of \cite{AmeMa:wav}, we
also propose a simple quantum mechanical point of view inspired by
Penrose's spin network theorem \cite{Pen}. In our case we
construct minimum uncertainty states $|j,\theta,\phi\>$ for each
spin $j$ in which expectations $\<f(x)\>$ behave approximately
like classical functions in polar coordinates $r,\theta,\phi$ with
$r=\lambda j$. In effect we view $\R^3_\lambda$ as a collection of
fuzzy spheres for all spins $j$ taken together. There are some similarities
also with the star product pointed out to us in \cite{HLS}.  

Finally, whereas the above includes electromagnetic theory on
$\RL$, we explain now that exactly this noncommutative space is
needed for a geometric picture underlying the approach to $2+1$
quantum gravity of \cite{bais,schroers}. When a Euclidean
signature and vanishing cosmological constant is assumed, the
gauge group of the classical gravitational action (as a
Chern-Simons field theory) is the group $ISO(3)$\cite{witten}.
Considering the three dimensional space as the direct product
$\Sigma \times \R$, where $\Sigma$ is Riemann surface of genus $g$
one can find the space of solutions of the gravitational field in
terms of the topology of $\Sigma$ \cite{alekseev,fock}. The
simplest case is to consider $\Sigma$ as a sphere with a puncture,
which represents the topological theory of one particle coupled to
gravity, it is known that the quantum states of this kind of
theory correspond to irreducible representations of the quantum
double $D(U(\su))$ \cite{bais}. A more detailed explanation, based
on representation theory, on how the quantum double is a
deformation ``quantization'' of the Euclidean group in three
dimensions can be found in \cite{schroers}. However, the direct
geometrical role of the quantum double has been missing except as
an `approximate' isometry of $\R^3$. Our present results therefore
provide a new point of view, namely of the quantum double symmetry
as an {\em exact} symmetry but of the noncommutative space $\RL$
on which we should build a noncommutative Chern-Simons action etc.
This fits with the discussion above that noncommutative spacetime
could be used as a better effective description of corrections to
geometry coming out of quantum gravity. Details of the required
noncommutative Chern-Simons theory as well as gravity in the frame
bundle approach of \cite{majid2} will be presented in a sequel.

\section{Mathematical Preliminaries}

Here we outline some notions from quantum group theory into which our
example fits. For Hopf algebras (i.e. quantum groups) we use the conventions of
\cite{majid}. It means an algebra $H$ equipped with a coproduct
$\Delta : H \rightarrow H \otimes H$, counit $\eps:H\to \C$ and
antipode $S:H\to H$. We will sometimes use the formal sum notation
$\Delta (a) = \sum a_{(1)} \otimes a_{(2)}$, for any $a\in H$. The
usual universal enveloping algebra algebra $\U$ has a structure of
cocommutative Hopf algebra generated by $1$ and $J_a$, $a=1,2,3$
with relations (\ref{J}) and \bray \Delta (J_a ) = J_a \otimes 1
+1 \otimes J_a ,\quad \epsilon (J_a)=0 ,\quad  S(J_a ) = -J_a.
\eray

We also recall that as for Abelian groups, for each Hopf algebra
there is a dual one where the product of one is adjoint to the
coproduct of the other. $\U$ is dually paired with the commutative
Hopf algebra $\CSU$ generated by coordinate functions ${t^i}_j$,
for $i,j=1,2$ on $SU(2)$ satisfying the determinant relation
${t^1}_1 {t^2}_2 -{t^1}_2 {t^2}_1 =1$ and with: \bray \Delta
({t^i}_j )= \sum_{k=1}^{2} {t^i}_k \otimes {t^k}_j ,\quad \epsilon
({t^i}_j ) = {\delta^i}_j,\quad S{t^i}_j  = {t^i}_j{}^{-1}. \eray
where inversion is as an algebra-valued matrix. The pairing
between the algebras $\U$ and $\CSU$ is defined by
\[
\langle \xi, f \rangle =  \frac{d}{dt} f(e^{t\xi} ) \mid_{t=0} ,
\]
where $\xi\in \su$ and $f\in \CSU$ which results in particular in:
\eqn{pairJt}{ \langle J_a , {t^i}_j \rangle = \frac{1}{2}
{{\sigma_a}^i}_j,} where ${{\sigma_a}^i}_j$ are the $i,j$ entries
of the Pauli matrices for $a=1\ldots 3$. We omit here a discussion
of unitarity here but this is implicit and achieved by making the
above into Hopf $*$-algebras, see \cite{majid} for further
details.

We also need standard notions of actions and coactions. A left
coaction of a Hopf algebra $H$ on a space $V$ means a map $V\to
H\tens V$ obeying axioms like those of an action but reversing all
maps. So a coaction of $\CSU$ essentially corresponds to an action
of $\U$ via the pairing. Examples are:
\begin{equation}
\label{leftadaction} \mbox{Ad}_L (h) (g) =h\trr g= \sum h_{(1)} g
S(h_{(2)} ).
\end{equation}
the left adjoint action \\ $\Ad_L:H\tens H\to H$. Its arrow-reversal
is the left adjoint coaction $\mbox{Ad}^L :H \rightarrow H\otimes
H$,
\begin{equation}
\label{leftcoaction} \mbox{Ad}^L (h) = \sum h_{(1)} S(h_{(3)})
\otimes h_{(2)}.
\end{equation}
There are also the regular action (given by the
product), regular coaction (given by \\ $\Delta:H\to H\tens H$), and
coadjoint actions and coregular actions of the dual, given via the
pairing from the adjoint and regular coactions,
etc.\cite{majid}. We will need the left coadjoint action of $H$
on a dual quantum group $A$:
\begin{equation}
\label{leftcoadaction}
\Ad_L^*(h)(\phi)= h\la \phi=\sum \phi_{(2)}
\<(S\phi_{(1)})\phi_{(2)},h\>,\quad \forall h\in H,\quad \phi\in
A
\end{equation}
and the right coregular action of $A$ on $H$ which we will view as
a left action of the opposite algebra $A^{\rm op}$:
\begin{equation}
\label{rightcoregaction} \phi\la h=\sum
\<\phi,h_{(1)}\>h_{(2)},\quad \forall h\in H,\quad \phi\in A.
\end{equation}

Given a quantum group $H$ dual to a quantum group $A$, there
is a quantum double written loosely as $D(H)$ and containing $H,A$
as sub-Hopf algebras. More precisely it is a double cross product
$A^{\rm op}\bowtie H$ where there are cross relations given by
mutual coadjoint actions\cite{majid}. Also, $D(H)$ is formally
quasitriangular  in the sense of a formal `universal R matrix'
$\CR$ with terms in $D(H)\tens D(H)$. The detailed structure of
$D(\U)$ is covered in Section~3 and in this case is more simply a semidirect
product $\CSU\crossed \U$ by the coadjoint action.

We will also need the quantum double $D(H)$ when $H$ is some other
quasitriangular quantum group such as $U_q(\su)$. This is a
standard deformation of (\ref{J}) and the coproduct etc. with a
parameter $q$. In this case there is a second `braided' or
covariantized version of $\C_q(SU(2))$ which we denote by $\BSU$.
Then
\eqn{Dbos}{D(U_q(\su))\isom \BSU\brcrossed U_q(\su)} where the
product is a semidirect one by the adjoint action of $U_q(\su)$
and the coproduct is also a semidirect one. We will use this
non-standard bosonisation version of $D(H)$ when $H$ is
quasitriangular. Also when $H$ is quasitriangular with $\CR$
nondegenerate, there is a third `twisting' version of the quantum
double:
\eqn{Dtwi}{ D(U_q(\su))\isom U_q(\su)\dbcocrossed_{\CR}U_q(\su)}
where the algebra is a tensor product and the coproduct is
\[
\Delta (h\otimes g )= {\CR}^{-1}_{23} \Delta_{H\otimes H}
(h\otimes g) {\CR}_{23}.
\]
We will use both versions in Section~7. Note that both
isomorphisms are formal but the right hand sides are well defined
and we take them as definitions. Especially, the isomorphism
(\ref{Dtwi}) is highly singular as $q\to 1$. In that limit the
twisted version tends to $U({so(1,3)})$ while the bosonisation
version tends to $U(iso(3))$.

Finally, we will need the notion of differential calculus on an
algebra $H$. This is common to several approaches to
noncommutative geometry including that of Connes \cite{Connes}. A
first order calculus means to specify $(\Omega^1,\extd)$, where
$\Omega^1$ is an $H-H$-bimodule, $\extd:H\to \Omega^1$ obeys the
Leibniz rule, \eqn{leibniz}{ \extd(hg) =(\extd h)g + h(\extd g).}
and $\Omega^1$ is spanned by elements of the form $(\extd h)g$. A
bimodule just means that one can multiply `1-forms' in $\Omega^1$
by `functions' in $H$ from the left or the right without caring
about brackets.

When we have a Hopf algebra $H$, a differential calculus can be
asked to be `bicovariant'\cite{woronowicz} which means that there
are left and right coactions of $H$ in $\Omega^1$ (a bicomodule)
which are themselves bimodule homomorphisms, and $\extd$
intertwines the coactions with the regular coactions of $H$ on
itself. Given a bicovariant calculus one can find invariant forms
\eqn{domega}{ \omega (h) = \sum (\extd h_{(1)})  S h_{(2)}} for
any $h\in H$. The span of such invariant forms is a space
$\Lambda^1$ and all of $\Omega^1$ can be reconstructed from them
via \eqn{omegad}{ \extd h=\sum \omega(h_{(1)}) h_{(2)}.} As a
result, the construction of a differential structure on a quantum
group rests on that of $\Lambda^1$, with $\Omega^1=\Lambda^1.H$.
They in turn can be constructed in the form
\[ \Lambda^1=\ker\eps/\CI\]
where $\CI\subset\ker\eps$ is some left ideal in $H$ that is
$\Ad^L$-stable \cite{woronowicz}. We will use this method in
Section~4 to introduce a reasonable calculus on $\U$. Some general
remarks (but not our calculus, which seems to be new) appeared in
\cite{majid2}.

Any bicovariant calculus has a `minimal' extension to an entire
exterior algebra\cite{woronowicz}. One uses the universal
R-matrix of the quantum double to define a braiding operator 
on $\Lambda^1\tens\Lambda^1$ and uses it to `antisymmetrize' the formal
algebra generated by the invariant forms. These and elements of
$H$ define $\Omega$ in each degree. In our case of $\U$, because
it is cocommutative, the braiding is the usual flip. Hence we have
the usual anticommutation relations among invariant forms. We also
extend $\extd:\Omega^k \rightarrow \Omega^{k+1}$ as a
(right-handed) super-derivation by
\[
\extd (\omega \wedge \eta ) = \omega \wedge \extd\eta +
(-1)^{\mbox{deg}\eta} \extd\omega \wedge \eta .
\]

A differential calculus is said to be inner if the exterior
differentiation in $\Omega^1$ (and hence in all degrees) is given
by the (graded) commutator with an invariant 1-form $\theta \in
\Lambda^1$, that is
\[
\extd\omega = \omega \wedge \theta - (-1)^{\mbox{deg}\omega}
\theta \wedge \omega .
\]
Almost all noncommutative geometries that one encounters are
inner, which is the fundamental reason that they are in many ways
better behaved than the classical case.

\section{The Quantum Double as Exact Isometries of $\RL$}

In this section we first of all recall the structure of the
quantum double $D(\U)$ in the context of Hopf
algebra theory. We will then explain its canonical action on a
second copy $\RL\isom \U$ arising from the general Hopf algebra
theory, thereby presenting it explicitly as an exact quantum
symmetry group of that. Here $x_a=\lambda J_a$ is the isomorphism
valid for $\lambda\ne 0$. By an exact quantum symmetry we mean
that the quantum group acts on $\RL$ with the product of $\RL$ an
intertwiner (i.e. the algebra is covariant).

Because $\U$ is cocommutative, its quantum double $\DU$ is a usual
crossed product \cite{majid}
\[
\DU =\CSU{}_{\Ad_L^*} \crossed \U
\]
where the action is induced by the adjoint action (it is the
coadjoint action on $\CSU$).  This crossed product is isomorphic
as a vector space with $\CSU \otimes \U$ but with algebra
structure given by
\[
(a\otimes h)(b\otimes g) =\sum a \Ad_L^*{}_{h_{(1)}} (b) \otimes
h_{(2)} g,
\]
for $a,b\in\CSU$ and $h,g\in\U$. In terms of the generators, the
left coadjoint action (\ref{leftcoadaction}) takes the form
\begin{equation}
\label{coadjoint} \Ad_L^*{}_{J_a} ({t^i}_j ) = \sum {t^k}_l
\langle J_a , S({t^i}_k ) {t^l}_j \rangle={1\over 2}\left(
t^i{}_k\sigma_a^k{}_l-\sigma_a^i{}_k t^k{}_j\right).
\end{equation}
As a result we find that $D(\U)$ is generated by $\U$ and $\CSU$
with cross relations \eqn{Jtcomm}{ [J_a, t^i{}_j]={1\over 2}\left(
t^i{}_k\sigma_a^k{}_j-\sigma_a^i{}_k t^k{}_j\right).} Meanwhile
the coproducts are the same as those of $\U$ and $\CSU$.

Next, a general feature of any quantum double is a canonical or
`Schr\"{o}dinger' representation, where $\U \subset \DU$ acts on
$\U$ by the left adjoint action (\ref{leftadaction}) and
$\CSU\subset D(\U)$ acts by the coregular one
(\ref{rightcoregaction}), see\cite{majid}. We denote the
acted-upon copy by $\RL$. Then $J_a$ simply act by
\begin{equation}
\label{lambdaadjoint}
J_a\trr f(x) =\lambda^{-1} \sum {x_a}_{(1)} f(x) S({x_a}_{(2)} )
=\lambda^{-1}[x_a,h],\quad \forall f(x)\in\RL
\end{equation}
e.g. \[ J_a \la x_a=\imath\eps_{abc}x_c\] while the co-regular
action reads
\[
t^i{}_j\la f(x) = \<t^i{}_j, f(x)_{(1)} \>f(x)_{(2)},\quad {\rm
e.g.}\quad t^i{}_j\la x_a=\frac{\l}{2}\sigma_a{}^i{}_j
1+\delta^i{}_j x_a.\] The general expression is given by a shuffle
product (see Section~4). {\em With this action, $\RL$ turns into a
left $\DU$-covariant algebra.}

In order to analyse the classical limit of this action, let us
consider the role of the numerical parameter $\lambda$ used to
define the algebra $\RL$. Considering the relations (\ref{x}) we
have already explained that $\RL$ becomes the usual algebra of
functions on $\R^3$ as $\lambda\to 0$. The same parameter
$\lambda$ can be introduced into the quantum double by means of a
redefinition of the generators of $\CSU$ to
\begin{equation}
\label{mgenerators}
{M^i}_j =\frac{1}{\lambda} \left( {t^i}_j -{\delta^i}_j \right) ,
\end{equation}
so that ${t^i}_j ={\delta^i}_j +\lambda {M^i}_j$.
We stress that we are dealing  with the same Hopf Algebra $\DU$,
but written in terms of new generators, it is only a change of
variables. The homomorphism property of $\Delta$ gives
\[
\Delta {M^i}_j = \sum_{k=1}^{2} \left( {\delta^i}_k \otimes
{M^k}_j + {M^i}_k \otimes {\delta^k}_j +\lambda {M^i}_k \otimes
{M^k}_j \right),
\]
while the condition on the determinant, ${t^1}_1 {t^2}_2 -{t^1}_2
{t^2}_1 =1$, implies that
\[
\mbox{Tr} (M)= {M^1}_1 +{M^2}_2 =-\lambda \mbox{det} (M).
\]
This means that in the limit $\lambda \rightarrow 0$, the elements
${M^i}_j$ have obey ${M^1}_1 =- {M^2}_2$ and $\CSU$ becomes the
commutative Hopf algebra $U (\R^3 )$. To make this explicit, we
can define the momentum generators
\begin{equation}
P_1 = -\imath \left( {M^1}_2 +{M^2}_1 \right)
,\quad P_2 =  {M^1}_2 -{M^2}_1 ,\quad  P_3 =
-\imath \left( {M^1}_1 - {M^2}_2 \right)
\end{equation}
or \eqn{momentum}{P_a=\sigma_a^i{}_j M^j{}_i,\quad a=1,2,3} (sum
over $i,j$). The inverse of this relationship is \eqn{MP}{
M^i{}_j=\frac{\imath}{2} \sigma_a{}^i{}_j P_a+{1\over
2}\delta^i{}_j P_0,\quad P_0={\rm Tr}(M)=-{2\over\lambda}
\left(1-\sqrt{1- {\lambda^2\over 4}P^2}\right).} The other square
root is also allowed but then $P_0$ is not $\CO(\lambda)$, i.e.
this is not the `patch' of $\CSU$ that concerns us. Note also that
there are unitarity conditions that we do not explicitly discuss
(if we put them in then the $P_a$ are hermitian). In these terms
we have
\[
\Delta P_a=P_a\tens 1+1\tens P_a+\CO(\lambda)
\]
so that we have the usual additive coproduct in the $\lambda\to 0$
limit. Meanwhile, the left coadjoint action (\ref{coadjoint})  and
the resulting cross relations in the double become
\[
\Ad_L^*{}_{J_a} (P_b ) =\imath\eps_{abc} P_c,\quad
[J_a,P_b]=\imath\eps_{abc}P_c.
\]
i.e. $D(\U)$ in the limit $\lambda\to 0$ with these generators
becomes the usual $U(iso(3))$. This part is essentially
known\cite{bais,schroers}.

Moreover, our action of these scaled generators on $\RL$ is:
\begin{equation}
\label{lambdacoregular}  M^i{}_j\la f(x)=\del^i{}_j(f(x)),\quad
{\rm e.g.}\quad   {M^i}_j\la x_a= \langle J_a ,{t^i}_j \rangle 1=
\frac{1}{2} {{\sigma_a}^i}_j 1
\end{equation}
where the operators $\del^i{}_j$ are the same as those in the next
section. We can also write the action of $P_a$ as partial
derivatives defined there (in (\ref{delijdela})) by
\[
P_a \trr f(x) = -\imath \partial_a f(x) \quad ,
\quad P_0 \trr f(x) = \frac{1}{c} \partial_0 f(x),
\]
where the constant $c$ is put in order to make the equations have
the same form as the classical ones, interpreting roughly the
$0$-direction as a ``time'' direction. This relation will become
clearer in section 7.

In the limit $\lambda\rightarrow 0$, the action of $J_a$
becomes usual rotations in three dimensional Euclidean space while
the action of $P_a$ becomes the action of translation operators of
the algebra $U(\R^3 )$, so we indeed recover the classical action
of $U(iso(3))$ on $\R^3$.   In three dimensional gravity,
considering the dimension of the gravitational constant $G_3$ and
the speed of light to be equal to 1, we have that $\lambda$ must
be proportional to the Planck constant \cite{schroers}.

Next, there are several applications of the action of the double
based on the above point of view. First and foremost, we could
look for a wave operator from a Fourier transform point of view as
in \cite{AmeMa:wav} (we give a different point of view later).
Namely we look for a Casimir of $D(\U)$ lying in momentum space
$\CSU$, and define the wave operator as its action. The possible
such Casimirs are the $\U$-invariant functions, which means
basically the class functions on $SU(2)$. In our case this just
means any function of the trace function $\tau=t^1{}_1+t^2{}_2$. 
The one suggested by the noncommutative geometry in the
next sections is
\eqn{waveP}{ \CE\equiv - P^2-{4\over \l^2}
\left(1-\sqrt{1-{\l^2\over 4} P^2}\right)^2={4\over\lambda}(\tau-2)}
and its action on $\RL$ is then the wave operator $\square$ on
degree zero in Section~5. Note that $S\tau=\tau$ for $\CSU$ so
any such wave operator is invariant under group inversion, which
appears as the antipode $SP_a=-P_a$.

A different question we can also ask is about the noncommutative
analogues of spherical harmonics as functions in $\RL$ in the
sense of irreducible representations $Y_l{}^m$ under the above
action (\ref{lambdaadjoint} of the rotation group. We find the
(unnormalized) lowest ones for $l\in {\mathbb Z}_+$ and $m=-l,
-l+1, \ldots , l$ as \bray
Y_0{}^0 &=&1 , \nonumber \\
Y_1{}^{\pm 1} &=& \mp \frac{1}{\sqrt{2}} (x_1 \pm \imath x_2 )
,\qquad Y_1{}^0 =  x_3 , \nonumber \\
Y_2{}^{\pm 2} &=&  (x_1 \pm \imath x_2 )^2
,\qquad Y_2{}^{\pm 1} = \mp  \left( (x_1 \pm \imath x_2 ) x_3 +
x_3 (x_1 \pm \imath x_2 ) \right) , \nonumber\\
Y_2{}^0 &=& \frac{1}{\sqrt{6}} \left( 4x_3^2 - (x_1 +\imath x_2
)(x_1 -\imath x_2 ) - (x_1 -\imath x_2 )(x_1 +\imath x_2 )
\right).\nonumber \eray   Let us note that such spherical
harmonics can have many  applications beyond their usual role in
physics. For example they classify the possible noncommutative
differential calculi on the classical coordinate algebra $\CSU$ 
which is dual to the space we study here.

\section{The 4-Dimensional Calculus on $\RL$}

The purpose of this section is to construct a bicovariant
calculus on the algebra $\UL$ following the steps outlined in Section
2, the calculus we obtain being that on the
algebra $\U$ on setting $\l =1$. We write $\RL$ as generated by
$x_+$, $x_-$ and $h$, say, and with the Hopf algebra structure
given explicitly in terms of the generators as
\bray
\label{lambdacommutations2}
[h, x_\pm]=\pm 2 \l x_\pm \quad ;\quad
[x_+ , x_- ]= \l h
\eray
and the additive coproduct as before. The
particular form of the coproduct, the relations and
(\ref{domega}) shows that $\extd\xi=\omega (\xi)$ for all $\xi\in
{\mathfrak su} (2)$. Because of the cocommutativity, all ideals in
$\RL$ are invariant under adjoint coactions (\ref{leftcoaction})
so that first order differential calculi $\Omega^1$ on $\RL$ are
classified simply by the ideals $\CI\subset \ker \epsilon$. In
order to construct an ideal of $\ker\epsilon$, consider a two
dimensional representation $\rho : \UL \rightarrow \End\C^2$,
which in the basis $\{ e_1 , e_2 \}$ of $\C^2$ is given by
\[
\begin{array}{ll}
\rho (x_+ ) e_1 =0 ,& \quad \rho (x_+ ) e_2 = \l e_1 ,\\
\rho (x_- ) e_1 = \l e_2 ,& \quad \rho(x_-) e_2 = 0 ,\\
\rho (h ) e_1 = \l e_1 ,& \quad \rho (h ) e_2 = - \l e_2 .
\end{array}
\]
The representation $\rho$ is a surjective map onto $M_2 (\C)$,
even when restricted to $\ker\epsilon$. The kernel of
$\rho\mid_{\ker\epsilon}$  is a 2-sided ideal in $\ker\epsilon$.
Then we have \eqn{quotient}{ M_2 (\C) \equiv \ker\epsilon / \ker
\rho .} This isomorphism allows us to identify the basic 1-forms
with $2\times 2$ matrices, $\{ e_{ij} \}$, for $i,j=1,2$, where $e_{ij}$ 
is the matrix with $1$ in the $(i,j)$ entry and $0$ otherwise. Then
the first order differential calculus is
\[
\Omega^1(\UL )= M_2 (\C ) \otimes \UL .
\]
The exterior derivative operator is
\[
\extd f(x)= \lambda^{-1}\sum \rho (f(x)_{(1)}-\epsilon
(f(x)_{(1)})1) f(x)_{(2)} =e_{ij} \del^i{}_j(f)
\]
where the last equality is a definition of the partial derivatives $\del^i{}_j:\RL\to
\RL$. In particular, we have \[ \extd
\xi=\lambda^{-1}\rho(\xi),\quad \forall \xi\in \su,\] which, along
with $\id$, span the whole space $M_2(\C)$ of invariant 1-forms. For a
general monomial $\xi_1 \ldots \xi_n$, the expression of the
derivative is
\[
\extd(\xi_1 \ldots \xi_n ) =\lambda^{-1}\sum_{k=1}^{n}
\sum_{\sigma \in S_{(n,k)}} \rho (\xi_{\sigma (1)} \ldots
\xi_{\sigma (k)} ) \xi_{\sigma (k+1)} \ldots \xi_{\sigma (n)} ,
\]
where $\sigma$ is a permutation of $1\ldots n$, such that $\sigma
(1) < \ldots <\sigma (k)$ and $\sigma (k+1) < \ldots <\sigma (n)$.
This kind of permutation is called a $(n,k)$-shuffle. And finally,
for a (formal power series) group-like element $g$ (where $\Delta
g=g\tens g$), the derivative is
\[
\extd g= \lambda^{-1}(\rho (g) -\theta )g.
\]

On our basis we have
\[
\extd x_+ = e_{12} , \quad \extd x_- =e_{21} , \quad \extd h =
e_{11} -e_{22}, \quad \theta = e_{11} +e_{22}.
\]
The compatibility conditions of this definition of the derivative
with the Leibniz rule is due to the following commutation relations between
the generators of the algebra and the basic 1-forms:
\bray
\label{formrelations1}
x_\pm \extd x_\pm &=& (\extd x_\pm) x_\pm ,\nonumber\\
x_\pm \extd x_\mp &=& (\extd x_\mp) x_\pm +\frac{\l}{2}
\left( \theta \pm\extd h \right) ,\nonumber\\
x_\pm \extd h &=& (\extd h x_\pm) \mp \l \extd x_\pm ,\nonumber\\
h \extd x_\pm &=& (\extd x_\pm) h \pm\l \extd x_\pm ,\nonumber\\
h \extd h &=& (\extd h) h + \l \theta , \nonumber\\
x_\pm \theta &=& \theta x_\pm +\l \extd x_\pm , \nonumber\\
h \theta &=& \theta h + \l \extd h.
\eray

From these commutation relations, we can see that this calculus is
inner, that is, the derivatives of any element of the algebra can
be basically obtained by the commutator with the 1-form
$\theta$. In the classical limit, this calculus turns out to be
the commutative calculus on usual three dimensional Euclidean space.
The explicit expression for the derivative of a general monomial
$x_-^a h^b x_+^c$ is given by
\bray
\label{derivative1}
\extd(x_-^a h^b x_+^c ) &=& \extd h
\left( \sum_{i=0}^{\left[\frac{b-1}{2} \right] } \left(
\begin{array}{c}
b\\
{2i+1}
\end{array}
\right)
\l^{2i} x_-^a h^{b-2i-1} x_+^c  \right) \nonumber\\
&+& \theta \left( \sum_{i=1}^{\left[\frac{b}{2} \right] } \left(
\begin{array}{c}
b\\
2i
\end{array}
\right)
\l^{2i-1} x_-^a h^{b-2i} x_+^c  \right) \nonumber\\
&+& \extd x_+ \left( \sum_{i=0}^{b} \left(
\begin{array}{c}
b\\
i
\end{array}
\right) \l^{i} c
x_-^a h^{b-i} x_+^{c-1}  \right) \nonumber\\
&+& \extd x_- \left( \sum_{i=0}^{b} \left(
\begin{array}{c}
b\\
i
\end{array}
\right) \l^{i}  a
x_-^{a-1} h^{b-i} x_+^{c}  \right) \nonumber\\
&+& \frac{1}{2} (\theta -\extd h )  \left( \sum_{i=0}^{b} \left(
\begin{array}{c}
b\\
i
\end{array}
\right) \l^{i+1} ac x_-^{a-1} h^{b-i} x_+^{c-1}  \right) , \eray
where the symbol $[z]$ denotes the greatest integer less than $z$
and only terms with $\ge 0$ powers of the generators included.
Note that this expression becomes in the limit $\l \rightarrow 0$
the usual expression for the derivative of a monomial in three
commuting coordinates.

In terms of the generators $x_a$ , $a=1\ldots 3$, which are
related to the previous generators by
\[
x_1 =\frac{1}{2} \left( x_+ +x_- \right) ,\quad x_2 =\frac{\imath}{2}
\left( x_- -x_+ \right) , \quad x_3 =\frac{1}{2} h ,
\]
we have
\eqn{Dxa}{ \extd x_a={1\over 2} \sigma_a,\quad
\theta=\sigma_0}
i.e. the Pauli matrices are nothing
other than three of our basic 1-forms, and together with
$\sigma_0=\id$ form a basis of the invariant 1-forms. The
commutation relations (\ref{formrelations1}) have a simple
expression:
\bray
\label{formrelations2}
x_a \extd x_b &=& (\extd x_b) x_a
+\frac{\imath\l}{2} \epsilon_{abc} \extd x_c
+\frac{\l}{4} \delta_{ab} \theta , \nonumber\\
x_a \theta &=& \theta x_a + \l \extd x_a .
\eray
In this basis the
partial derivatives defined by
\eqn{dela}{ \extd f(x)=(\extd
x_a)\del^a f(x)+\theta \frac{1}{c} \del^0 f(x)}
are related to the previous ones by
\eqn{delijdela}{ \del^i{}_j={1\over
2}\sigma_a{}^i{}_j\del^a+{1\over 2c}\sigma_0{}^i{}_j \del^0}
as in (\ref{MP}).
The exterior derivative of a general monomial
$x_1^a x_2^b x_3^c$ is quite complicated to write down explicitly
but we find it as:
\bray
\label{derivative2}
&\, & \extd(x_1^a x_2^b
x_3^c ) = \sum_{i=0}^{\left[ \frac{a}{2}\right]}
\sum_{j=0}^{\left[ \frac{b}{2}\right]} \sum_{k=0}^{\left[
\frac{c}{2}\right]} \theta \frac{\l^{2(i+j+k)-1}}{2^{2(i+j+k)}}
 \left(
\begin{array}{c}
a\\
2i
\end{array}
\right)
 \left(
\begin{array}{c}
b\\
2j
\end{array}
\right)
 \left(
\begin{array}{c}
c\\
2k
\end{array}
\right)
x_1^{a-2i} x_2^{b-2j} x_3^{c-2k} \nonumber\\
&+& \sum_{i=0}^{\left[ \frac{a}{2}\right]} \sum_{j=0}^{\left[
\frac{b}{2}\right]} \sum_{k=0}^{\left[ \frac{c-1}{2}\right]} \extd
x_3 \frac{\l^{2(i+j+k)}}{2^{2(i+j+k)}}
 \left(
\begin{array}{c}
a\\
2i
\end{array}
\right)
 \left(
\begin{array}{c}
b\\
2j
\end{array}
\right)
 \left(
\begin{array}{c}
c\\
2k+1
\end{array}
\right)
x_1^{a-2i} x_2^{b-2j} x_3^{c-2k-1} \nonumber\\
&+& \sum_{i=0}^{\left[ \frac{a}{2}\right]} \sum_{j=0}^{\left[
\frac{b-1}{2}\right]} \sum_{k=0}^{\left[ \frac{c}{2}\right]} \extd
x_2 \frac{\l^{2(i+j+k)}}{2^{2(i+j+k)}}
 \left(
\begin{array}{c}
a\\
2i
\end{array}
\right)
 \left(
\begin{array}{c}
b\\
2j+1
\end{array}
\right)
 \left(
\begin{array}{c}
c\\
2k
\end{array}
\right)
x_1^{a-2i} x_2^{b-2j-1} x_3^{c-2k} \nonumber\\
&+& \sum_{i=0}^{\left[ \frac{a}{2}\right]} \sum_{j=0}^{\left[
\frac{b-1}{2}\right]} \sum_{k=0}^{\left[ \frac{c-1}{2}\right]}
\imath \extd x_1 \frac{\l^{2(i+j+k)+1}}{2^{2(i+j+k)+1}}
 \left(
\begin{array}{c}
a\\
2i
\end{array}
\right)
 \left(
\begin{array}{c}
b\\
2j+1
\end{array}
\right)
 \left(
\begin{array}{c}
c\\
2k+1
\end{array}
\right)
x_1^{a-2i} x_2^{b-2j-1} x_3^{c-2k-1} \nonumber\\
&+& \sum_{i=0}^{\left[ \frac{a-1}{2}\right]} \sum_{j=0}^{\left[
\frac{b}{2}\right]} \sum_{k=0}^{\left[ \frac{c}{2}\right]} \extd
x_1 \frac{\l^{2(i+j+k)}}{2^{2(i+j+k)}}
 \left(
\begin{array}{c}
a\\
2i+1
\end{array}
\right)
 \left(
\begin{array}{c}
b\\
2j
\end{array}
\right)
 \left(
\begin{array}{c}
c\\
2k
\end{array}
\right)
x_1^{a-2i-1} x_2^{b-2j} x_3^{c-2k} \nonumber\\
&-& \sum_{i=0}^{\left[ \frac{a-1}{2}\right]} \sum_{j=0}^{\left[
\frac{b}{2}\right]} \sum_{k=0}^{\left[ \frac{c-1}{2}\right]}
\imath \extd x_2 \frac{\l^{2(i+j+k)+1}}{2^{2(i+j+k)+1}}
 \left(
\begin{array}{c}
a\\
2i+1
\end{array}
\right)
 \left(
\begin{array}{c}
b\\
2j
\end{array}
\right)
 \left(
\begin{array}{c}
c\\
2k+1
\end{array}
\right)
x_1^{a-2i-1} x_2^{b-2j} x_3^{c-2k-1} \nonumber\\
&+& \sum_{i=0}^{\left[ \frac{a-1}{2}\right]} \sum_{j=0}^{\left[
\frac{b-1}{2}\right]} \sum_{k=0}^{\left[ \frac{c}{2}\right]}
\imath \extd x_3 \frac{\l^{2(i+j+k)+1}}{2^{2(i+j+k)+1}}
 \left(
\begin{array}{c}
a\\
2i+1
\end{array}
\right)
 \left(
\begin{array}{c}
b\\
2j+1
\end{array}
\right)
 \left(
\begin{array}{c}
c\\
2k
\end{array}
\right)
x_1^{a-2i-1} x_2^{b-2j-1} x_3^{c-2k} \nonumber\\
&+& \sum_{i=0}^{\left[ \frac{a-1}{2}\right]} \sum_{j=0}^{\left[
\frac{b-1}{2}\right]} \sum_{k=0}^{\left[ \frac{c-1}{2}\right]}
\theta \frac{\l^{2(i+j+k)+2}}{2^{2(i+j+k)+3}}
 \left(
\begin{array}{c}
a\\
2i+1
\end{array}
\right)
 \left(
\begin{array}{c}
b\\
2j+1
\end{array}
\right)
 \left(
\begin{array}{c}
c\\
2k+1
\end{array}
\right) x_1^{a-2i-1} x_2^{b-2j-1} x_3^{c-2k-1}  \nonumber\\
&-& \frac{\theta}{\l} x_1^a x_2^b x_3^c.
\eray

In both cases the expression for the derivatives of plane waves is
very simple. In terms of generators $x_a$, the derivative of the
plane wave $e^{\imath\sum_a k^a x_a }=e^{i k\cdot x}$ is given by
\bray \label{planewave1} \extd e^{\imath k\cdot x} =\left\{
\frac{\theta}{\l} \left( \cos \left( \frac{\l \mid
k\mid}{2}\right)  -1 \right) +   \frac{2\imath\sin \left( \frac{\l
\mid k\mid }{2}\right)}{\l \mid k\mid } k\cdot \extd\, x \right\}
e^{\imath k\cdot x} . \eray One can see that the limit $\l
\rightarrow 0$ gives the correct formula for the derivative of
plane waves, that is
\[
\lim_{\l \rightarrow 0} \extd e^{\imath k\cdot x} = \left(
\sum_{a=1}^3 \imath k_a \extd \bar x_a \right) e^{\imath k\cdot
\bar x}=\imath k\cdot (\extd \bar x)e^{\imath k\cdot \bar x},
\]
where at $\lambda=0$ on the right hand side we have the classical
coordinates and the classical 1-forms in usual three
dimensional commutative calculus. In terms of the generators
$x_{\pm}$, $h$, the plane wave $e^{\imath(k_+ x_+ +k_- x_- +k_0
h)} =e^{\imath k\cdot x}$ is given by \bray \label{planewave2}
\extd e^{\imath k\cdot x} &=& \left\{ \frac{\theta}{\l} \left(
\cos \left( \l \sqrt{k_0^2 +k_+ k_- } \right)  -1 \right) \right.
\nonumber\\
&&+ \left. \frac{\imath(k_+ \extd x_+ +k_- \extd x_- +k_0 \extd
h)}{\l \sqrt{k_0^2 +k_+ k_- }} \left( \sin \left( \l \sqrt{k_0^2
+k_+ k_- } \right)\right) \right\} e^{ik\cdot x} . \eray

This calculus is four dimensional, in the sense that one has four
basic 1-forms, but these dimensions are entangled in a nontrivial
way. For example, note that they satisfy the relation
\[
\epsilon_{abc} x_a (\extd x_b) x_c =0.
\]
We can see that in the classical limit $\l \rightarrow 0$, the
calculus turns out to be commutative and the extra dimension,
namely the one-dimensional subspace generated by the 1-form
$\theta$, decouples totally from the calculus generated by the
other three 1-forms. The relation between this extra dimension and
quantization can also be perceived by considering the derivative
of the Casimir operator
\[
C = \sum_{a=1}^3  (x_a )^2 ,
\]
which implies
\[
\extd C = 2 \sum_{a=1}^3  (\extd x_a) x_a +\frac{3\l}{4} \theta .
\]
The coefficient of the term in $\theta$ is exactly the eigenvalue
of the Casimir in the spin $\frac{1}{2}$ representation, the same
used to construct the differential calculus, and also vanishes
when $\l \rightarrow 0$. We shall see later that this extra
dimension can also be seen as a remnant of the time
coordinate in the q-Minkowski space $\R^{1,3}_q$ when the
limit $q\rightarrow 1$ is taken. A semi-classical analysis on this
calculus can also be made in order to recover an interpretation of
time in the three dimensional noncommutative space.

We can also construct the full exterior algebra $\Omega^{\cdot}
(\UL ) = \bigoplus_{n=0}^{\infty} \Omega^n (\UL)$. In our case the
general braiding\cite{woronowicz} becomes the trivial flip
homomorphism because the right invariant basic 1-forms are also
left invariant. Hence our basic 1-forms in $M_2(\C)$ are totally
anticommutative and their usual antisymmetric wedge product
generates the usual exterior algebra on the vector space $M_2 (\C
)$. The full $\Omega^\cdot(\RL)$ is generated by these and
elements of $\RL$ with the relations (\ref{formrelations1}). The
exterior differentiation in $\Omega^{\cdot} (\UL )$ is given by
the graded commutator with the basic 1-form $\theta$, that is
\[
\extd\omega = \omega \wedge \theta - (-1)^{\mbox{deg}\omega}
\theta \wedge \omega.
\]
In particular, the basic 1-forms $M_2(\C)$ are all closed, among
which $\theta$ is not exact. The cohomologies of this calculus
were also calculated giving the following results:

\begin{theorem} The noncommutative de Rham cohomology of $\RL$ is
\[
H^0 =\C.1,\quad H^1 =\C.\theta, \quad  H^2 = H^3 = H^4 = \{ 0\} .
\]
\end{theorem}
\proof This is by direct (and rather long) computation of the
closed forms and the exact ones in each degree using the
explicit formula (\ref{derivative1}) on general monomials. To
give an example of the procedure, we will do it in some detail
for the case of 1-forms. Take a general 1-form
\[
\omega =\alpha (\extd x_+) x_-^a h^b x_+^c + \beta (\extd x_-)
x_-^m h^n x_+^p + \gamma (\extd h) x_-^r h^s x_+^t + \delta \theta
x_-^u h^v x_+^w ,
\]
and impose $\extd\omega =0$. We start analysing the simplest
cases, and then going to more complex ones.

Taking $\beta =\gamma =\delta =0$, then
\[
\omega =\alpha (\extd x_+) x_-^a h^b x_+^c .
\]
The term in $dx_- \wedge dx_+$ leads to the conclusion that $c=0$.
Similarly, the term in $\extd h\wedge \extd x_+$ leads to $b=0$ so that
\[
\omega =\alpha (\extd x_+) x_+^c =\extd\left( \frac{1}{c+1}
x_+^{c+1} \right) ,
\]
which is an exact form, hence belonging to the null cohomology class.
The cases $\alpha =\gamma =\delta =0$ and $\alpha =\beta =\delta =0$ 
also lead to exact forms. The case $\alpha =\beta =\gamma =0$ leads to
the 1-form
\[
\omega =\delta \theta x_-^u h^v x_+^w .
\]
The vanishing of the term in $\extd x_+ \wedge \theta$ implies that
$w=0$, the term in $\extd x_- \wedge \theta$ vanishes if and only
if $u=0$ and the term in $\extd h \wedge \theta$ has its vanishing
subject to the condition $v=0$. Hence we have only the closed, non-exact
form $\theta$ from this case.

Let us now analyse the case with two non zero terms:
\[
\omega =\alpha (\extd x_+) x_-^a h^b x_+^c + \beta (\extd x_-)
x_-^m h^n x_+^p .
\]
The vanishing condition in the term on $\extd x_- \wedge \extd
x_+$ reads
\[
\alpha \sum_{i=0}^b
\left( \begin{array}{c} b\\ i \ea \right) \l^i a
x_-^{a-1} h^{b-i} x_+^c = \beta \sum_{i=0}^n
\left( \begin{array}{c} n\\ i \ea \right) \l^i p
x_-^{m} h^{n-i} x_+^{p-1} .
\]
Then we conclude that $b=n$, $a-1 =m$, $c=p-1$ and $\alpha a
=\beta (c+1)$. The vanishing of the term in $\extd h\wedge \extd
x_+$ reads
\[
\sum_{i=0}^{\left[ \frac{b-1}{2} \right]}
\left( \begin{array}{c} b\\ 2i +1 \ea \right) \l^{2i}
x_-^{a} h^{b-2i-1} x_+^c = \frac{1}{2}\sum_{i=0}^b
\left( \begin{array}{c} b\\ i \ea \right) \l^{i+1} ac
x_-^{a-1} h^{b-i} x_+^{c-1} .
\]
The terms in odd powers of $\l$ vanish if and only if $ac=0$. Then
the left hand side vanishes if and only if $b=0$. The case $a=0$,
implies that $\beta =0$, which reduces to the previous case already
mentioned. For the case $c=0$ we have $\beta =\alpha a$ so that
\[
\omega = \alpha  \left( (\extd x_+) x_-^a + a (\extd x_-)
x_-^{a-1} x_+ \right) .
\]
It is easy to see that $\omega$ is closed if and only if $a=1$. But
\[
(\extd x_+) x_- + (\extd x_-) x_+ = \extd \left( x_- x_+ +
\frac{\l}{2} h \right) - \frac{\l}{2} \theta ,
\]
which is a form homologous to $\theta$. It is a long, but
straightforward calculation to prove that all the other cases of
closed 1-forms rely on these cases above mentioned.

The proof that all higher cohomologies are trivial is also an
exhaustive analysis of all the possible cases and inductions on
powers of $h$, as exemplified here for the 4-forms: It is clear
 that all 4-forms
\[
\omega = \extd x_- \wedge \extd h \wedge \extd x_+ \wedge \theta
x_-^m h^n x_+^p
\]
are closed. We use induction on $n$ to prove that there exists a
three form $\eta$ such that $\omega =d\eta$. For $n=0$, we have
\[
\extd x_- \wedge \extd h \wedge \extd x_+ \wedge \theta x_-^m
x_+^p = \extd\left( -\frac{1}{m+1} \extd h \wedge \extd x_+ \wedge
\theta x_-^{m+1} x_+^p \right).
\]
Suppose that there exist 3-forms $\eta_k$, for $0\leq k<n$, such
that
\[
\extd x_- \wedge \extd h \wedge \extd x_+ \wedge \theta x_-^m h^k
x_+^p = \extd\eta_k ,
\]
then \bray \extd x_- \wedge \extd h \wedge \extd x_+ \wedge \theta
x_-^m h^n x_+^p &=& \extd\left( -\frac{1}{m+1} \extd h \wedge
\extd x_+ \wedge \theta x_-^{m+1} h^n
x_+^p \right) \nonumber\\
&-& \extd x_- \wedge \extd h \wedge \extd x_+ \wedge \theta
\sum_{i=1}^n \left(
\begin{array}{c} n\\ i \ea \right) \l^{i}
x_-^m h^{n-i} x_+^p \nonumber\\
&=& \extd\left( -\frac{1}{m+1} \extd h \wedge \extd x_+ \wedge
\theta x_-^{m+1} h^n x_+^p - \sum_{i=1}^n \left( \begin{array}{c}
n\\ i \ea \right) \l^{i} \eta_{n-i} \right).\nonumber \eray Hence
all 4-forms are exact. The same procedure is used to show the
triviality of the other cohomologies. \eproof

For $\RL$ we should expect the cohomology to be trivial, since
this corresponds to Stokes theorem and many other aspects taken
for granted in physics. We find almost this except for the
generator $\theta$ which generates the calculus and which has no
3-dimensional classical analogue. We will see in Section~7 that
$\theta$ is a remnant of a time direction even though from the
point of view of $\RL$ there is no time coordinate. The cohomology
result says exactly that $\theta$ is an allowed direction but not
$\extd$ of anything.

\section{Hodge $*$-Operator and Electromagnetic Theory}

The above geometry also admits a metric structure. It is known
that any nondegenerate bilinear form $\eta \in \Lambda^1 \otimes
\Lambda^1$ defines an invariant metric on the Hopf algebra $H$
\cite{majid2}. For the case of $\RL$ we can define the metric
\eqn{metric}{\eta = \extd x_1 \otimes \extd x_1 + \extd x_2
\otimes \extd x_2 + \extd x_3 \otimes \extd x_3 + \mu \theta
\otimes \theta} for a parameter $\mu$. This bilinear form is
non-degenerate, invariant by left and right coactions and
symmetric in the sense that $\wedge (\eta) =0$. With this metric
structure, it is possible to define a Hodge $*$-operator and then
explore the properties of the Laplacian and find some physical
consequences. Our picture is similar to \cite{xavier1} where the
manifold is similarly three dimensional but there is an extra time
direction $\theta$ in the local cotangent space.

The Hodge $*$-operator on an $n$ dimensional calculus (for which
the top form is of order $n$), over a Hopf algebra $H$ with metric
$\eta$ is a map $* : \Omega^k \rightarrow \Omega^{n-k}$  given
by the expression
\[
* (\omega_{i_1} \ldots \omega_{i_k} ) = \frac{1}{(n-k)!}
\epsilon_{i_1 \ldots i_k i_{k+1} \ldots i_n} \eta^{i_{k+1} j_1}
\ldots \eta^{i_n j_{n-k}} \omega_{j_1} \ldots \omega_{j_{n-k}} .
\]
In the case of the algebra $\RL$, we have a four dimensional
calculus with $\omega_1 =\extd x_1$, $\omega_2 =\extd x_2$,
$\omega_3 =\extd x_3$, $\omega_4 =\theta$. The components of the
metric inverse, as we can see from (\ref{metric}), are $\eta^{11}
= \eta^{22} =\eta^{33} =1$, and $\eta^{44} =\frac{1}{\mu}$. The
arbitrary factor $\mu$ in the metric can be set by imposing
conditions on the map $*^2$. Then we have two possible choices for
the constant $\mu$: The first is $\mu =1$ making a four dimensional
Euclidean geometry, then for a $k$-form $\omega$ we have the
constraint $**(\omega) =(-1)^{k(4-k)} \omega$. The second
possibility is $\mu =-1$, then the metric is Minkowskian and the
constraint on a $k$-form $\omega$ is $**(\omega) =(-1)^{1+ k(4-k)}
\omega$. In what follows, we will be using the Minkowskian
convention on the grounds that this geometry on $\RL$ is a remnant
of a noncommutative geometry on a $q$-deformed version of the
Minkowski space, as we shall explain in Section~7. The expressions
for the Hodge $*$-operator are summarized as follows: \bray *1 &=&
-\extd x_1 \wedge \extd x_2 \wedge \extd x_3 \wedge \theta ,
\nonumber\\
\, * \extd x_1 &=& -\extd x_2 \wedge \extd x_3 \wedge \theta ,
\nonumber\\
\, * \extd x_2 &=& \extd x_1 \wedge \extd x_3 \wedge \theta ,
\nonumber\\
\, * \extd x_3 &=& -\extd x_1 \wedge \extd x_2 \wedge \theta ,
\nonumber\\
\, * \theta &=& -\extd x_1 \wedge \extd x_2 \wedge \extd x_3 ,
\nonumber \eray \bray  *(\extd x_1 \wedge \extd x_2 ) &=& -\extd
x_3 \wedge \theta ,
\nonumber\\
\, *(\extd x_1 \wedge \extd x_3 ) &=& \extd x_2 \wedge \theta ,
\nonumber\\
\, *(\extd x_1 \wedge \theta ) &=& \extd x_2 \wedge \extd x_3 ,
\nonumber\\
\, *(\extd x_2 \wedge \extd x_3 ) &=& -\extd x_1 \wedge \theta ,
\nonumber\\
\, *(\extd x_2 \wedge \theta ) &=& -\extd x_1 \wedge \extd x_3 ,
\nonumber\\
\, *(\extd x_3 \wedge \theta ) &=& \extd x_1 \wedge \extd x_2 ,
\nonumber \eray \bray  *(\extd x_1 \wedge \extd x_2 \wedge \extd
x_3 ) &=& -\theta ,
\nonumber\\
\, *(\extd x_1 \wedge \extd x_2 \wedge \theta ) &=& -\extd x_3 ,
\nonumber\\
\, *(\extd x_1 \wedge \extd x_3 \wedge \theta ) &=& \extd x_2 ,
\nonumber\\
\, *(\extd x_2 \wedge \extd x_3 \wedge \theta ) &=& -\extd x_1 ,
\nonumber\\
\, *(\extd x_1 \wedge \extd x_2 \wedge \extd x_3 \wedge \theta )
&=&1 .\label{hodge} \eray

Given the Hodge $*$-operator, one can write, for
example, the coderivative $\delta =*\extd*$ and the Laplacian
operator $\Delta =\delta \extd +\extd\delta$. Note that the
Laplacian maps to forms of the same degree. We prefer to work
actually with the `Maxwell-type' wave operator
\eqn{maxwave}{\square =\delta \extd =*\extd*\extd} which is just
the same on degree 0 and the same in degree 1 if we work in a
gauge where $\delta=0$. In the rest of this section, we are going
to describe some features of the electromagnetic theory arising in
this noncommutative context. The electromagnetic theory is the
analysis of solutions $A\in \Omega^1 (\UL)$ of the equation
$\square A= \mbox{J}$ where $\mbox{J}$ is a 1-form which can be
interpreted as a ``physical'' source.  We demonstrate the theory
on two natural choices of sources namely an electrostatic and a
magnetic one. We start with spin 0 and we limit ourselves to
algebraic plus plane wave solutions.

\subsection{Spin 0 modes}

The wave operator on $\Omega^0 (\RL) =\RL$ is computed from the
definitions above as
\[
\square=*\extd*\extd=(\del^a)^2-\frac{1}{c^2}(\del^0)^2
\]
where the partials are defined by (\ref{dela}). The algebraic
massless modes $\ker\square$ are given by
\begin{itemize}
\item Polynomials of degree one: $f(x)= \alpha +\beta_a x_a$.
\item Linear combinations of polynomials of the type
$f(x)= (x_a^2 -x_b^2 )$.
\item Linear combinations of quadratic monomials of the type,
$f(x)= \alpha_{ab}x_a x_b$, with $a\neq b$.
\item The three particular combinations $f(x)=x_1 x_2 x_3 -
\frac{\imath \l}{4} (x_a^2)$, for $a=1,2,3$.
\end{itemize}

General eigenfunctions of $\square$ in degree 0 are the plane
waves; the expression for their derivatives can be seen in
(\ref{planewave1}). Hence
\[
\square e^{\imath k\cdot x} =-\frac{1}{\l^2}\left\{ 4 \sin^2
\left( \frac{\l \mid k\mid}{2} \right) +\left( \cos \left(
\frac{\l \mid k\mid}{2} \right) -1\right)^2 \right\} e^{\imath
k\cdot x} .
\]
It is easy to see that this eigenvalue goes in the limit $\l
\rightarrow 1$ to the usual eigenvalue of the Laplacian in three
dimensional commutative space acting on plane waves.

\subsection{Spin 1 electromagnetic modes} On $\Omega^1 (\UL )$,
the Maxwell operator $\square_1=*\extd*\extd$ can likewise be
computed explicitly. If we write $A=(\extd x_a)A^a+\theta A^0$ for
functions $A_\mu$, then
\[ F=\extd A=\extd x_a\wedge \extd x_b \del^b A^a+\extd
x_a\wedge\theta \frac{1}{c}\del^0 A^a
+\theta\wedge \extd x_a \del^a A^0
\]
If we break this up into electric and magnetic parts in the usual
way then
\[
B_a=\eps_{abc}\del^b A^c,\quad E_a=\frac{1}{c} \del^0 A^a-
\del^a A^0.\]
These computations have just the same form as for usual spacetime.
The algebraic zero modes $\ker \square_1$ are given by

\begin{itemize}
\item Forms of the type $A = \extd x_a (\alpha +\beta_a x_a +
\gamma_a x_a^2 )$ with curvature
\[ F= \frac{\l}{4} \gamma_a \extd x_a \wedge \theta.\]
\item Forms of the type $A = \beta_{ab} (\extd x_a) x_b$, with
$a\neq b$ and curvature 
\[ F= \beta_{ab} \extd x_a \wedge \extd x_b, \quad a\neq b.\]
\item Forms of the type $A =\theta f$ with $f\in \ker \square$. 
The curvatures for the latter three $f(x)$ shown above are
\bray
 F&=& -2 (\extd x_a \wedge \theta x_a - \extd x_b \wedge \theta x_b
) ,
\nonumber\\
F&=&\alpha_{ab}\left(\theta\wedge(\extd x_a)x_b+\theta\wedge 
(\extd x_b)x_a+{\imath\lambda\over 2}\eps_{abc}\theta
\wedge\extd x_c\right)\nonumber\\
F&=& -\extd x_1 \wedge \theta \left( x_2 x_3 + \frac{\imath \l}{2}
x_1 \right) -\extd x_2 \wedge \theta \left( x_1 x_3 - \frac{\imath
\l}{2} x_2 \right) -\extd x_3 \wedge \theta \left( x_1 x_2 +
\frac{\imath \l}{2} x_3 \right)
\nonumber\\
& &\quad - \frac{\imath \l}{2} \extd x_a \wedge\theta x_a\nonumber,
\quad a=1,2,3. \eray
\end{itemize}
These are `self-propagating' electromagnetic modes or solutions of
the sourceless Maxwell equations for a 1-form or `gauge potential'
$A$.

\subsection{Electrostatic solution}
Here we take a uniform source in the `purely time' direction
$\mbox{J}=\theta$. In this case the solution of the gauge
potential is
\[
A=\frac{1}{6} \theta C,
\]
where $C$ is the Casimir operator. The curvature operator, which
in this case can be interpreted as an electric field is given by
\[
F=\extd A=\frac{1}{3} \left( \theta \wedge (\extd x_1) x_1 +
\theta \wedge (\extd x_2) x_2 + \theta \wedge (\extd x_3) x_3
\right) .
\]
If $\theta$ is viewed as a
time direction then this curvature is a radial electric field. It has
field strength increasing with the radius, which is a kind of solution
exhibiting a confinement behaviour. 

\subsection{Magnetic solution}  Here we take
a uniform electric current density along a direction vector 
$k\in\R^3$, i.e. $\mbox{J}= k\cdot \extd x=\sum_a k^a\extd x_a$. 
In this case, the gauge potential can be written as
\[
A=\frac{1}{4} \left\{ \left( \sum_{a=1}^3 k_a \extd x_a \right) C
+ \frac{\theta}{2} \left( \sum_{a=1}^3 k_a x_a \right) -
\sum_{a=1}^3 k_a (\extd x_a) x_a^2 \right\} .
\]
The field strength is: \bray F=\extd A &=& \frac{1}{2} \left\{
\extd x_1 \wedge \extd x_2 (k_1 x_2 -k_2 x_1 )
+ \extd x_1 \wedge \extd x_3 (k_1 x_3 -k_3 x_1 ) +\right. \nonumber\\
&&+ \left. \extd x_2 \wedge \extd x_3 (k_2 x_3 -k_3 x_2 ) \right\}
. \eray If we decompose the curvature in the usual way then this
is an magnetic field in a direction $k\times x$ (the vector cross
product). This is a `confining' (in the sense of increasing with
normal distance) version of the field due to a current in
direction $k$.

We have considered for the electromagnetic solutions only uniform 
sources $J$; we can clearly put in a functional dependence for the
coefficients of the source to similarly obtain other solutions of
both the electric and magnetic types. Solutions more similar to
the usual decaying ones, however, will not be polynomial (one
would need the inverse of $\sum_a x_a^2$) and are therefore well
outside our present scope; even at a formal level the problem of
computing $\extd (\sum_a x_a^2)^{-1}$ in a closed form appears to
be formidable. On the other hand these matters could probably be
addressed by completing to $C^*$-algebras and using the functional
calculus for such algebras.

\subsection{Spin $1/2$ equation} For completeness, let us mention
here also a natural spin 1/2 wave operator, namely the Dirac
operator. We consider the simplest (Weyl) spinors as two
components $\psi^i\in \RL$.  In view of the fact that the partial
derivatives $\del^i{}_j$ already form a matrix, and following the
similar phenomenon  as for quantum groups \cite{majid2}, we are
led to define \eqn{dirac}{ (\dsl\psi)^i=\del^i{}_j\psi^j.}
According to (\ref{delijdela}) this could also be written as
\[
\dsl={1\over 2}\sigma_a\del^a+{1\over 2c}\del^0
\]
where the second
term is suggested by the geometry over an above what we might also
guess. This term is optional in the same way as $(\del^0)^2$ in
$\square$ is not forced by covariance, and is $\CO(\lambda)$ for
bounded spatial derivatives.

Here $\dsl$ is covariant under the quantum double action in
Section~3 as follows (the same applies without the $\del^0$ term).
The action of $J_a$ on $\RL$ is that of orbital angular momentum
and we have checked already that $\square$ on degree 0 is
covariant. For spin $\h$ the total spin should be \eqn{spinhalf}{
S_a={1\over 2}\sigma_a+J_a} and we check that this commutes with
$\dsl$:
\begin{eqnarray*}
(S_a\dsl\psi)^i&=&{1\over 2}\sigma_a{}^i{}_j\del^j{}_k\psi^k+
J_a M^i{}_j\la\psi^j\\
&&={1\over
2}\sigma_a{}^i{}_jM^j{}_k\la\psi^k+[J_a,M^i{}_j]\la\psi^j+M^i{}_j
J_a\la \psi^j\\
&&={1\over 2} M^i{}_j\la \sigma_a{}^j{}_k\psi^k+M^i{}_j J_a\la
\psi^j=M^i{}_j \la (S_a\psi)^j=(\dsl S_a\psi)^i\end{eqnarray*}
where we used the relations (\ref{Jtcomm}) (those with $M^i{}_j$
have the same form) and the action (\ref{lambdacoregular}). The
operator $\dsl$ is clearly also translation invariant under $\CSU$
since the $\del^i{}_j$ mutually commute. The operators $\sigma_a$
and $\del^i{}_j$ also commute since one acts on the spinor indices
and the other on $\RL$, so $S_a$ in place of $J_a$ still gives a
representation of $D(\U)$ on spinors, under which $\dsl$ is
covariant.

\subsection{Yang-Mills $U(1)$ fields}

Finally, also for completeness, we mention that there is a
different $U(1)$ theory which behaves more like Yang-Mills. Namely
instead of $F=\extd A$ as in the Maxwell theory, we define
$F=\extd A+A\wedge A$ for a 1-form $A$. This transforms by
conjugation as $A\mapsto g A g^{-1}+g\extd g^{-1}$ and is a
nonlinear version of the above, where $g\in \RL$ is any invertible
element, e.g. a plane wave. In this context one would
expect to be able to solve for zero-curvature, i.e. $A$ such that
$F(A)=0$ and thereby demonstrate the Bohm-Aharanov effect etc.
This is part of the nonlinear theory, however, and beyond our
present scope.

\section{Differential Calculus on the Quantum Sphere}

In this section we briefly analyse what happens if we try to set
the `length' function given by the Casimir $C$ of $\RL$ to a fixed
number, i.e. a sphere. We take this at unit radius, i.e. we define
$S^2_\lambda$ as the algebra  $\RL$ with the additional relation
\eqn{sph}{ C\equiv\sum_{a=1}^{3} \left( x_a \right)^2 =1. } This
immediately gives a `quantisation condition' for the constant $\l$
if the algebra is to have an irreducible representation, namely
$\l = \frac{1}{\sqrt{j(j+1)}}$ for some $j\in \frac{1}{2}\Z_+$.
The image of $S_\lambda^2$ in such a spin $j$ representation is a
$(2j+1)\times (2j+1)$-matrix algebra which can be identified with
the class of noncommutative spaces known as `fuzzy spheres'
\cite{madore,ARS,bal,pinzul,ramgoolan}. In these works one does
elements of noncommutative differential geometry directly on
matrix algebras motivated by thinking about them as a projection
of $\U$ in the spin $j$ representation and the greater the spin
$j\rightarrow \infty$, the greater the resemblance with a
classical sphere. The role of this in our case is played by $\l
\rightarrow 0$ according to the above formula. On the other hand
note that we are working directly on $S^2_\lambda$ and are not
required to look in one or any irreducible representation, i.e.
this is a slightly a more geometrical approach to `fuzzy spheres'
where we deform the conventional geometry of $S^2$ by a parameter
$\lambda$ and do not work with matrix algebras.

Specifically, when we make the constraint (\ref{sph}), the four
dimensional calculus given by relations (\ref{formrelations2}) is
reduced to a three dimensional calculus on the sphere because
\[
\extd C = \sum_{a=1}^{3}  2 (\extd x_a) x_a +\frac{3\l}{4} \theta
=0 ,
\]
which means that $\theta$ can be written as an expression on
$\extd x_a$. The remaining relations are given by \bray
\label{lsphere1} x_a \extd x_b &=& (\extd x_b) x_a +\frac{i}{2} \l
\epsilon_{abc} \extd x_c
-\frac{2}{3}  \delta_{ab} \sum_{d=1}^{3} (\extd x_d) x_d , \nonumber\\
\l^2 \extd x_a &=& \frac{4i}{3} \l  \epsilon_{abc} (\extd x_b) x_c
- \frac{16}{9} \sum_{d=1}^{3} (\extd x_d) x_d x_a . \eray In the
limit $\l \rightarrow 0$ we recover the ordinary two dimensional
calculus on the sphere, given in terms of the classical variables
$\bar{x}_a =\lim_{\lambda \rightarrow 0} x_a$. This can be seen by
the relation
\[
\sum_{a=1}^{3} ( \extd{\bar{x}}_a) \bar{x}_a =0 ,
\]
allowing to write one of the three 1-forms in terms of the other
two. For example, in the region where $\bar{x}_3 =
\sqrt{1-{\bar{x}}_1^2 -{\bar{x}}_2^2}$ is invertible, one can
write
\[
\extd{\bar{x}}_3 = -\frac{\bar{x}_1}{\sqrt{1-{\bar{x}}_1^2 -
{\bar{x}}_2^2 }} \extd{\bar{x}}_1
-\frac{\bar{x}_2}{\sqrt{1-{\bar{x}}_1^2 -{\bar{x}}_2^2 }}
\extd{\bar{x}}_2 .
\]

\section{The Space $\UL$ as a Limit of $q$-Minkowski Space}

In this section, we will express the noncommutative space $\UL$ as
a spacelike surface of constant time in a certain scaling limit of
the standard  $q$-deformed Minkowski space $\R^{1,3}_q$ in
\cite{CWSS,majidlorentz,majid}. This is defined in
\cite{majid} as the algebra of $2\times 2$ braided (Hermitian)
matrices $BM_q (2)$ generated by 1 and
\[
\mbox{\bf u} =\left( \begin{array}{cc}
                                      a & b \\
                                      c & d
                     \end{array} \right) ,
\]
with the commutation relations \cite{majid}
\bray \label{braided1}
ba &=& q^2 ab ,\nonumber\\
ca &=& q^{-2} ac , \nonumber \\
da &=& ad , \nonumber \\
bc &=& cb + (1- q^{-2})a(d-a) ,\nonumber \\
db &=& bd + (1- q^{-2})ab ,\nonumber \\
cd &=& dc + (1- q^{-2})ca . \eray
If we choose a suitable set of
generators, namely \cite{majidlorentz}
\[
\tilde{t} =\frac{qd+q^{-1} a}{2} , \quad \tilde{x} =\frac{b +c}{2}
, \quad \tilde{y} =\frac{b -c}{2i} , \quad \tilde{z} =\frac{d
-a}{2}
\]
then the braided determinant \eqn{bdet}{\underline{\mbox{det}}
(\mbox{\bf u}) =ad -q^2 cb}
can be written as
\[
\underline{\mbox{det}} (\mbox{\bf u}) = \frac{4q^2}{(q^2 +1)^2}
\tilde{t}^2 - q^2 \tilde{x}^2 -q^2 \tilde{y}^2 - \frac{2(q^4 +1)
q^2}{(q^2 +1)^2} \tilde{z}^2 +2q \left( \frac{q^2 -1}{q^2 +1}
\right)^2 \tilde{t} \tilde{z} .
\]
This expression, in the limit $q\rightarrow 1$ becomes the usual
Minkowskian metric on $\R^{1,3}$. Here we will consider a
different scaled limit related to the role of this algebras as
braided enveloping algebra of a braided Lie algebra 
$\widetilde{{\frak su}_q(2)}$, see \cite{xavier2} for
a recent treatment. This is such that we can still have a
noncommutative space even when $q\rightarrow 1$. Defining new
generators
\begin{equation}
\label{newgenerators} x_+ =\frac{c}{(q-q^{-1})},\quad x_-
=\frac{b}{(q-q^{-1})},\quad h =\frac{a -d}{(q-q^{-1})},\quad t =
\frac{ qd +q^{-1} a}{c (q+q^{-1})},
\end{equation}
and considering the commutation relations (\ref{braided1}), we
have \bray \label{braided2} [x_+ ,x_- ] &=& q^{-1} c t h +
q^{-1} \frac{(q-q^{-1})}{(q+q^{-1})} h^2 ,\nonumber\\
q^{-2} hx_+ &=& x_+ h + q^{-2} (q+q^{-1})c  x_+ t , \nonumber\\
q^{2} hx_- &=& x_- h - (q+q^{-1})c  x_- t , \nonumber\\
t x_{\pm} &=& x_{\pm} t , \nonumber\\
t h &=& ht . \eray In the limit $q\rightarrow 1$, we obtain the
commutation relations \eqn{Rc}{ [x_a,x_b]=\imath c t\,
\eps_{abc}x_c,\quad [x_a,t]=0} of the so-called homogenized
universal enveloping algebra $\widetilde{\U}$, which we will
denote by $\R_c^{1,3}$. Here $c$ is a parameter required by
dimensional analysis (of dimension $m\, s^{-1}$). When $ct=\lambda $
we recover exactly the relations (\ref{lambdacommutations2}) of
$\RL$. So the noncommutative space that we have studied in
previous sections is the `slice' at a certain time of
$\R_c^{1,3}$, which in turn is a contraction of $\R_q^{1,3}$. The
possibility of these two $q\to 1$ limits where one give a
classical coordinate algebra and the other gives essentially its
dual (an enveloping algebra) is called a `quantum-geometry duality
transformation`.

We now go further and also obtain the differential structure on
$\RL$ via this scaling limit. Thus, the algebra $\R_q^{1,3}=BM_q
(2)$ has a standard $U_q(\su)$-covariant noncommutative
differential calculus whose commutation relations between basic
1-forms and the generators of the algebra, are given by
\cite{majid} \bray \label{braided3}
a\extd a &=& q^2 (\extd a) a ,\nonumber\\
a\extd b &=& (\extd b) a ,\nonumber\\
a\extd c &=& q^2 (\extd c) a +(q^2 -1) (\extd a) c ,\nonumber\\
a\extd d &=& (\extd d) a +(q^2 -1) (\extd b) c +
(q-q^{-1})^2 (\extd a) a ,\nonumber\\
b\extd a &=& q^2 (\extd a) b + (q^2 -1) (\extd b) a ,\nonumber\\
b\extd b &=& q^2 (\extd b) b ,\nonumber\\
b\extd c &=& (\extd c) b +(1-q^{-2} ) ((\extd d) a +(\extd a) d )
+
(q-q^{-1})^2 (\extd b) c -(2-3q^{-2} +q^{-4} ) (\extd a) a,\nonumber\\
b\extd d &=& (\extd d) b +(q^2 -1) (\extd b) d +(q^{-2} -1)(\extd
b) a
+(q-q^{-1})^2 (\extd a) b ,\nonumber\\
c\extd a &=& (\extd a) c ,\nonumber\\
c\extd b &=& (\extd b) c +(1-q^{-2} ) (\extd a) a ,\nonumber\\
c\extd c &=& q^2 (\extd c) c ,\nonumber\\
c\extd d &=& q^2 (\extd d) c +(q^2 -1) (\extd c) a ,\nonumber\\
d\extd a &=& (\extd a) d +(q^2 -1) (\extd b) c
+(q-q^{-1})^2 (\extd a) a ,\nonumber\\
d\extd b &=& q^2 (\extd b) d +(q^2 -1) (\extd a) b ,\nonumber\\
d\extd c &=& (\extd c) d +(q^2 -1) (\extd d) c
+(q-q^{-1})^2 (\extd c) a +(q^{-2} -1) (\extd a) c ,\nonumber\\
d\extd d &=& q^2 (\extd d) d +(q^2 -1) (\extd c) b +(q^{-2} -1)
(\extd b) c -(1-q^{-2} )^2 (\extd a) a. \eray

This is designed in the $q\to 1$ limit to give the usual
commutative calculus on classical $\R^{1,3}$. In order to obtain a
noncommutative calculus in our noncommutative scaled limit
$q\rightarrow 1$, we have to also redefine the derivative operator
by a scale factor
\[
\dd = (q-q^{-1}) \extd .
\]
This scaled derivative gives the following expressions for the
basic 1-forms: \bray
\dd x_+ &=& \extd c = (q-q^{-1}) \extd x_+ ,\nonumber\\
\dd x_- &=& \extd b = (q-q^{-1}) \extd x_- ,\nonumber\\
\dd h &=& \extd a -\extd d  = (q-q^{-1}) \extd h .\nonumber
\eray
Define also the basic 1-form
\[
\theta = q\extd d +q^{-1} \extd a ,
\]
which allows us to write
\begin{equation}
\label{dt} \dd t = \frac{(q-q^{-1})}{c (q+q^{-1})} \theta .
\end{equation}
This new set of generators and basic 1-forms satisfy the following
relations:
\bray
\label{braided4}
x_+ \dd x_+ &=& q^2 (\dd x_+) x_+ ,\nonumber\\
x_+ \dd x_- &=& (\dd x_-) x_+ +\frac{q^{-1}}{(q+q^{-1})} \theta c
t
+\frac{1}{(q+q^{-1})} (\dd h) ct +\CO (q-q^{-1})  ,\nonumber\\
x_+ \dd h &=& (\dd h) x_+ -q\dd x_+ +\CO (q-q^{-1}) ,\nonumber\\
x_- \dd x_+ &=& (\dd x_+) x_- +\frac{q^{-3}}{(q+q^{-1})} \theta ct
-\frac{(2-q^{-2} )}{(q+q^{-1})} (\dd h) ct+\CO (q-q^{-1}),\nonumber\\
x_- \dd x_- &=& q^2 (\dd x_-) x_-  ,\nonumber\\
x_- \dd h &=& q^2 (\dd h) x_- + q^{-1} (\dd x_-) ct +\CO
(q-q^{-1})
,\nonumber\\
h \dd x_+ &=& (\dd x_+ )h + q  (\dd x_+) ct+\CO (q-q^{-1}),\nonumber\\
h \dd x_- &=& (\dd x_- h) - q  (\dd x_-) ct+\CO (q-q^{-1}),\nonumber\\
h \dd h &=& (\dd h) h +\frac{2q}{(q+q^{-1})}  \theta c t
+\CO (q-q^{-1}) , \nonumber\\
x_+ \theta &=& \theta x_+ + q^2  (\dd x_+) c t +\CO (q-q^{-1}) ,
\nonumber\\
x_- \theta &=& \theta x_- + q^2 (\dd x_-) c t  +\CO (q-q^{-1}) ,
\nonumber\\
h \theta &=& \theta h +\frac{2q}{(q+q^{-1})}  (\dd h) c t +
\CO (q-q^{-1}) ,\nonumber\\
 t \dd x_+ &=& (\dd x_+)  t +\CO (q-q^{-1}) ,\nonumber\\
 t \dd x_- &=& (\dd x_-)  t +\CO (q-q^{-1}) ,\nonumber\\
 t \dd h &=& (\dd h)  t +\CO (q-q^{-1}) ,\nonumber\\
 t \theta &=& \theta  t +\CO (q-q^{-1}) . \eray In the limit
$q\rightarrow 1$ we recover the relations (\ref{formrelations1})
by setting $ct=\lambda$. Then the calculus on $\UL$ can be seen as
the pull-back to the time-slice of the scaled limit of the
calculus on $q$-deformed Minkowski space. Unlike for usual $\R^3$,
the $\extd t$ direction in our noncommutative case does not
`decouple' and has remnant $\theta$. In other words, {\em the geometry
of $\RL$ remembers that it is the pull-back of a relativistic
theory}.

Finally, let us recall the action of the $q$-Lorentz group on the
$\R_q^{1,3}$ and analyse its scaled limit when $q\rightarrow 1$.
The appropriate $q$-Lorentz group can be written as the double cross
coproduct $\UQ\dbcocrossed\UQ$. The Hopf algebra $\UQ$ is the standard
q-deformed Hopf algebra which we write explicitly as 
generated by $1$, $X_+$, $X_-$ and $q^{\pm
\frac{H}{2}}$ with \bray \label{uqsl2} && q^{\pm \frac{H}{2}}
X_{\pm} q^{\mp
\frac{H}{2}} = q^{\pm 1} X_{\pm}, \qquad [X_+ ,X_- ] =
\frac{q^H -q^{-H}}{q-q^{-1}} ,\nonumber\\
&& \Delta (X_{\pm}) = X_\pm\tens q^{H\over 2}+q^{-{H\over 2}}\tens
X_\pm, \qquad \Delta (q^{\pm \frac{H}{2}} ) =
q^{\pm \frac{H}{2}} \otimes q^{\pm \frac{H}{2}} , \nonumber \\
&&\epsilon( X_{\pm} ) =0, \qquad\qquad \epsilon (q^{\pm \frac{H}{2}} ) =1
,
\nonumber\\
&&S(X_{\pm}) = -q^{\pm 1}X_\pm, \qquad S( q^{\pm \frac{H}{2}} ) =
q^{\mp \frac{H}{2}} . \eray It is well-known that one may also
work with these generators in an $R$-matrix form
\eqn{mlmatrices}{
\L^+ = \left(
\begin{array}{cc}
      q^{\frac{H}{2}}           &    0\\
q^{-\frac{1}{2}} (q-q^{-1}) X_+ &  q^{-\frac{H}{2}} \ea
\right),\quad \L^- = \left(
\begin{array}{cc}
q^{-\frac{H}{2}} & q^{\frac{1}{2}} (q^{-1} -q) X_-  \\
        0        &  q^{\frac{H}{2}}
\ea \right)}
and most formulae are usually expressed in terms of
these matrices of generators. In particular, the $q$-Lorentz group
has two mutually commuting copies of $\UQ$, so let us denote the
generators of the first copy by $\M^\pm$ or $Y_\pm,G$ (related as
for $\L^\pm$ and $X_\pm, H$ in (\ref{mlmatrices})) and the
generators of the second copy of $\UQ$ by $\N^\pm$ or $Z_\pm,T$
(similarly related). The actions on $\R_q^{1,3}$ are given in
\cite{majid} in an R-matrix form
\begin{equation}
\label{braction} \N^{\pm}{}^k{}_l \trr {\u^i}_j = \langle
\N^{\pm}{}^k{}_l , {t^m}_j \rangle {\u^i}_m,\quad \M^{\pm}{}^k{}_l
\trr {\u^i}_j = \langle S\M^{\pm}{}^k{}_l, {t^i}_m \rangle
{\u^m}_j.
\end{equation}
Here $\langle S\M^{\pm}{}^k{}_l , {t^i}_j \rangle$ and $\langle
\N^{\pm}{}^k{}_l , {t^i}_j \rangle$ are the $i,j$ matrix entries
of the relevant functions of $Y_\pm,G$ and $Z_\pm,T$ respectively
in the Pauli matrix representation (as in (\ref{pairJt}) in other
generators). We need the resulting actions more explicitly, and
compute them as:

\bray \label{ac1} \frac{q^G -q^{-G}}{q-q^{-1}}
\trr\begin{pmatrix}h&x_-\\ x_+& t\end{pmatrix} &=&\begin{pmatrix}
-\frac{2 ct}{q-q^{-1}} - \frac{q-q^{-1}}{q+q^{-1}} h &  -x_- \\
x_+&\frac{q-q^{-1}}{q+q^{-1}} t -
\frac{q-q^{-1}}{c (q+q^{-1})^2} h\end{pmatrix} \nonumber\\
Y_+ \trr\begin{pmatrix}h&x_-\\ x_+& t\end{pmatrix}
&=&\begin{pmatrix} -qx_+&   -\frac{q ct}{q-q^{-1}} +\frac{h}{q+q^{-1}} 
\\  0 & -\frac{q-q^{-1}}{c (q+q^{-1})} x_+
\end{pmatrix} \nonumber\\
Y_- \trr\begin{pmatrix}h&x_-\\ x_+& t\end{pmatrix}
&=&\begin{pmatrix}q^{-1}x_-& 0\\ -\frac{q^{-1} ct}{q-q^{-1}} -
\frac{h}{q+q^{-1}} & -\frac{q-q^{-1}}{c (q+q^{-1})} x_-
\end{pmatrix}. \eray
\bray \label{ac2} \frac{q^T -q^{-T}}{q-q^{-1}} \trr
\begin{pmatrix}h&x_-\\ x_+& t\end{pmatrix}
&=&\begin{pmatrix} \frac{2 ct}{q-q^{-1}} +
\frac{q-q^{-1}}{q+q^{-1}} h &  -x_- \\ x_+ &
-\frac{q-q^{-1}}{q+q^{-1}} t +
\frac{q-q^{-1}}{c (q+q^{-1})^2} h \end{pmatrix}\nonumber \\
Z_+ \trr
\begin{pmatrix}h&x_-\\ x_+& t\end{pmatrix}
&=&\begin{pmatrix}  -x_+ & \frac{ct}{q-q^{-1}}
+\frac{qh}{q+q^{-1}}
\\ 0&  \frac{q(q-q^{-1})}{c (q+q^{-1})} x_+ \end{pmatrix} \nonumber\\
Z_- \trr\begin{pmatrix}h&x_-\\ x_+& t\end{pmatrix}
&=&\begin{pmatrix}  x_- & 0  \\ \frac{ct}{q-q^{-1}} - \frac{q^{-1}
h}{q+q^{-1}} & \frac{q^{-1} (q-q^{-1})}{c (q+q^{-1})}
x_-\end{pmatrix}. \eray
We are now able to see that these actions
(\ref{ac1}) and (\ref{ac2}) blow up in the limit $q\rightarrow 1$
because of some singular terms appearing in their expressions.
Hence the scaling limit $\R_c^{1,3}$ is no longer Lorentz
invariant.

On the other hand we also have the same quantum group symmetry in
an isomorphic form
  $\BSU \brcrossed \UQ$ for $q\ne 1$, and this version survives.
The braided
algebra $\BSU$ here is simply the braided matrices $BM_q (2)$ with
the additional condition $\underline{\mbox{det}} (\u )=1$ (i.e.
geometrically, it is the mass-hyperboloid in $q$-Minkowski space).
To be clear, the generators of $\BSU$ in this crossed product will
be denoted by $\bar{\u}$ and the generators of $\UQ$ in this cross
product will be denoted by $\L^{\pm}$ or $X_\pm,H$ as before. The
isomorphism with the $q$-Lorentz group in the form above is given
by the assignments \cite{majid}:
\begin{equation}
\label{iso} \bar{\u}\tens 1 \mapsto \M^+ S(\M^- )\tens 1,  \quad
1\tens \L^{\pm} \mapsto \M^{\pm} \otimes \N^{\pm} .
\end{equation}
Under the isomorphism (\ref{iso}), the expressions (\ref{ac1}) and
(\ref{ac2}) become the action of \\
$\BSU \brcrossed \UQ$ on $BM_q (2)$ given by
\[
\bar{\u}\trr \u = \M^+ S(\M^- ) \trr  \u,  \quad \L^{\pm} \trr \u
= \M^{\pm} \trr (\N^{\pm} \trr \u ) .
\]
On the generators (\ref{newgenerators}) the action of $\BSU$ reads
\bray
\label{ac3}
{\bar{\u}^1}_1 \trr h &=& -ct +qh
-\frac{q(q-q^{-1})}{q+q^{-1}} h, \nonumber\\
{\bar{\u}^1}_1 \trr x_+ &=& qx_+ , \nonumber \\
{\bar{\u}^1}_1 \trr x_- &=& q^{-1} x_- , \nonumber \\
{\bar{\u}^1}_1 \trr t &=& \frac{q^2 +q^{-2}}{q+q^{-1}} t -
\frac{(q-q^{-1})^2}{c (q+q^{-1})^2 } h , \nonumber \\
{\bar{\u}^1}_2 \trr h &=& q^{-2} (q-q^{-1} ) x_- \nonumber\\
{\bar{\u}^1}_2 \trr x_+ &=& -q^{-2} c t -
\frac{q^{-1} (q-q^{-1})}{q+q^{-1}} h , \nonumber \\
{\bar{\u}^1}_2 \trr x_- &=& 0 , \nonumber \\
{\bar{\u}^1}_2 \trr t &=& -\frac{q-q^{-1}}{q+q^{-1}} t +
\frac{q^{-1} (q-q^{-1})^2}{c(q+q^{-1})^2} h\eray \bray
{\bar{\u}^2}_1 \trr h &=& - (q-q^{-1} ) x_+ \nonumber\\
{\bar{\u}^2}_1 \trr x_+ &=& 0, \nonumber \\
{\bar{\u}^2}_1 \trr x_- &=&  -ct +
\frac{q^{-1} (q-q^{-1})}{q+q^{-1}} h , \nonumber \\
{\bar{\u}^2}_1 \trr t &=& -\frac{q (q-q^{-1})}{q+q^{-1}} t +
\frac{(q-q^{-1})^2}{c (q+q^{-1})^2} h, \nonumber \\
{\bar{\u}^2}_2 \trr h &=& ct +qh - q^{-1} (q-q^{-1})ct -
\frac{q^{-1}(q-q^{-1}) -q^{-2} (q-q^{-1})^2}{q+q^{-1}} h,
\nonumber\\
{\bar{\u}^2}_2 \trr x_+ &=& q^{-1} x_+ + q^{-1} (q-q^{-1})^2 x_+ ,
\nonumber \\
{\bar{\u}^2}_2 \trr x_- &=& q x_- , \nonumber \\
{\bar{\u}^2}_2 \trr t &=& \frac{2t}{q+q^{-1}}
+\frac{(q-q^{-1})^2}{q+q^{-1}} t + \frac{(q-q^{-1})^2 -q^{-1}
(q-q^{-1})^3}{c (q+q^{-1})^2 } h . \eray The action of $\UQ$
is given by

\bray \label{ac4} \frac{q^H -q^{-H}}{q-q^{-1}}
\trr\begin{pmatrix}h&x_-\\ x_+& t\end{pmatrix}
&=&\begin{pmatrix}0& -(q+q^{-1}) x_-\\   (q+q^{-1}) x_+ & 0\end{pmatrix} 
\nonumber\\
X_+ \trr \begin{pmatrix}h&x_-\\ x_+& t\end{pmatrix}
&=&\begin{pmatrix}  -q (q^{\frac{1}{2}} +q^{-\frac{1}{2}}) x_+
&q^{\frac{1}{2}} h
\\ 0 &  0\end{pmatrix}\nonumber\\
X_- \trr \begin{pmatrix}h&x_-\\ x_+& t\end{pmatrix}
&=&\begin{pmatrix} q^{-\frac{1}{2}} (q+q^{-1}) x_- & 0 \\ -
q^{-\frac{1}{2}} h & 0\end{pmatrix}. \eray In the limit
$q\rightarrow 1$, the crossed product $\BSU \brcrossed \UQ$
becomes the double $\DU=\CSU\crossed \U$ as studied in Section~3.
The elements ${{\bar{\u}}^i}_j$ become in the limit the ${t^i}_j$,
and $X_{\pm}$ and $H$ become the usual $\su$ generators equivalent
to the $J_a$ there. (More precisely, we should map $\bar u^i{}_j$
to $St^i{}_j$ for the action to become the right coregular one
which we viewed in Section~3 as a left coaction.) Finally, this
action of the double on $BM_q (2)$ thus becomes in the scaling
limit $q\to 1$ an action of $D(\U)$ on $\R_c^{1,3}$ in the form \[
[x_+,x_-]=2ct h,\quad [h,x_\pm]=\pm ctx_\pm\] with the same change
of variables to $x_a$ as in Section~4. The result is
\[ M^i{}_j\la t=0,\quad M^i{}_j\la x_a={ct\over
2\l}\sigma_a{}^i{}_j,\quad J_a\la t=0,\quad J_a\la
x_a=\imath\eps_{abc}x_c.\] This is consistent with the time slice
$ct=\lambda$ and gives the action of the quantum double in
Section~3 as in fact the nonsingular version of scaled limit of
the $q$-Lorentz symmetry on the $q$-Minkowski space.

One can also analyse a different time-slice of $\R_q^{1,3}$,
namely, the quotient obtained by imposing the condition $ct =q^2
+q^{-2} -1$. This algebra is the reduced braided algebra $BM_q
(2)^{\mbox{red}}$, see\cite{xavier2}, with commutation relations
\bray x_+ x_-  &=& x_- x_+ + q^{-1} (q^2 +q^{-2} -1) h +
\frac{(q-q^{-1})}{(q+q^{-1})} h^2 ,\nonumber\\
q^{-2} hx_+ &=& x_+ h + q^{-2} (q^2 +q^{-2} -1)(q+q^{-1}) x_+ ,
\nonumber\\
q^{2} hx_- &=& x_- h - (q^2 +q^{-2} -1)(q+q^{-1}) x_-.\nonumber
\eray This is also known in the literature as the `Witten algebra'
\cite{lebruyn,witten} and in a scaled limit $q\rightarrow 1$ it
likewise turns into the universal enveloping algebra $\U$. A
calculus on this reduced algebra, however, is not obtained from
the calculus given by relations (\ref{braided3}); consistency
conditions result in the vanishing of all derivatives $\extd a$,
$\extd b$ and $\extd c$ (note that the constraint on $t$ allows
one to write $d$ in terms of the other generators).

\section{Quantum mechanical interpretation and semiclassical limit
of $\RL$}

Finally, we turn to the important question of how to relate
expressions in the noncommutative geometry to ordinary numbers in
order to compare with experiment. We will first explain why a
normal ordering postulate as proposed in \cite{AmeMa:wav} is not
fully satisfactory and then turn to a quantum mechanical approach. 
Thus, one idea is to write elements of $\RL$ as $:f(x):$ where
$f(x_1,x_2,x_3)$ is a classical function defined by a powerseries
and $:\ :$ denotes normal ordering when we use noncommutative
variables $x_i$. If one sticks to this normal ordering one can use
it to compare classical with quantum expressions and express the
latter as a strict deformation of the former controlled by the
parameter $\lambda$ governing the noncommutativity  in
(\ref{lambdacommutations2}). This will extend to the rest of the
geometry and allows an order-by-order analysis. For example, the
noncommutative partial derivatives $\del_a$ defined in
(\ref{dela}) have the expressions to lowest order \bray
\label{semiclassicalpartial}
\partial_1 :f(x): &=&: {\bar{\partial}}_1 f(x): +\frac{\imath \l}{2}
{\bar{\partial}}_2 {\bar{\partial}}_3 f(x) , \nonumber\\
\partial_2: f(x): &=& :{\bar{\partial}}_2 f(x): -\frac{\imath
\l}{2}
{\bar{\partial}}_1 {\bar{\partial}}_3 f(x) , \nonumber\\
\partial_3: f(x) :&=&: {\bar{\partial}}_3 f(x): +\frac{\imath
\l}{2}
{\bar{\partial}}_2 {\bar{\partial}}_2 f(x) , \nonumber\\
\frac{1}{c} \partial_0 : f(x): &=& \frac{\l}{4} \left(
({\bar{\partial}}_1)^2 f(x) + ({\bar{\partial}}_2)^2 f(x) +
({\bar{\partial}}_3)^2 f(x) \right) , \eray where 
${\bar{\partial}}_a$ are the usual partial derivatives in
classical variables and we do not write the normal ordering on
expressions already $O(\lambda)$ since the error is higher order.
Note that the expression for $\frac{1}{c}
\partial_0$ is one order of $\l$ higher than the other partial
derivatives, which is another way see that this direction is an
anomalous dimension originating in the quantization process. The
physical problem here is that the normal ordering is somewhat
arbitrary; for algebras such as (\ref{kappamink}) or for usual
phase space, putting all  $t$ or $p$ to one side makes a degree of
sense physically, as well as mathematically because the algebra is
solvable. But in the simple case such as $\RL$, each of the
$x_1,x_2,x_3$ should be treated equally. Or one could use other
coordinates such as $x_-,h,x_+$ in keeping with the Lie algebra
structure, etc.; all different ordering schemes giving a different
form of the lowest order corrections and hence different
predictions. Choosing a natural ordering is certainly possible but
evidently would require further input into the model. 

On the other hand, we can take a more quantum mechanical line and
consider our algebra $\RL$ as, after all, a spin system. The main
result of this section is to introduce `approximately classical'
states' $|j,\theta,\phi\>$ for this system inspired in part by the
theorem of Penrose\cite{Pen} for spin networks, although not
directly related to that. Penrose considered networks labelled by
spins and showed how to assign probabilities to them and
conditions for when the network corresponds approximately to
spin measurements oriented with relative angles $\theta,\phi$.
In a similar spirit we consider the problem of reconstructing
classical angles form the noncommutative geometry.

We let  $V^{(j)}$ be the vector space which carries a unitary
irreducible representation of spin $j\in\frac{1}{2}{\mathbb Z}_+$,
generated by states $\mid j, m \rangle$, with $m=-j, \ldots , j$
such that \bray x_{\pm} \mid j, m \rangle &=& \l \sqrt{ (j\mp
m)(j\pm m+1)}
\mid j, m \pm 1 \rangle , \nonumber\\
h \mid j, m \rangle &=& 2\l m \mid j, m \rangle . \nonumber \eray
The projection of $\RL$ to an irreducible representation of spin
$j$ is geometrically equivalent to a restriction to a fuzzy
sphere \cite{bal,madore}, because the value of the Casimir
$x\cdot x$   is  $\l^2 j(j+1)$ in this representation. We have
discussed this in Section 6, where we set $x\cdot x=1$ and
considered the algebra geometrically as such a fuzzy sphere under
a quantisation condition for $\lambda$. By contrast in this
section we leave $x\cdot x$ unconstrained and consider the
geometry of our noncommutative three dimensional space $\RL$ as
the sum of geometries on all fuzzy spheres with the $V^{(j)}$
representation picking out the one of radius $\sim$ $\lambda j$.
Thus we use the Peter-Weyl decomposition of $\C(SU_2)$ into matrix
elements of irreducible representations regarded as functions on
$SU_2$, which gives (up to some technical issues about
completions) a similar decomposition for its dual as $\RL= \oplus_j
\End(V^{(j)}$. This also underlies the spherical harmonics in
Section~3.

Next, for each fixed spin $j$ representation we look for
normalised states $|j,\theta,\phi\>$ parameterized by $0\leq
\theta \leq \pi$ and $0\leq \varphi \leq 2\pi$, such that \bray \<
j, \theta , \varphi \mid x_1 |j, \theta , \varphi
\rangle &=&  r\sin \theta \cos \varphi , \nonumber\\
\<j, \theta , \varphi \mid x_2 | j, \theta , \varphi
\rangle &=&  r \sin \theta \sin \varphi , \nonumber\\
\< j, \theta , \varphi \mid x_3 |  j, \theta , \varphi \rangle &=&
 r \cos \theta  . \label{polar} \eray where $r$ is some constant
(independent of $\theta,\phi$) which we do not fix. Rather, in the
space of such states and possible $r\ge 0$, we seek to minimise
the normalised variance \eqn{varj}{ \delta ={\<x\cdot x\>-\<x\>\cdot
\<x\>\over \<x\>\cdot \<x\>}} where $\<\ \>=\<j,\theta,\phi|\ |j,
\theta,\phi\>$ is the expectation value in our state and we regard
$\<x_a\>$ as a classical vector in the dot product. Thus we seek
states which are `closest to classical'. This is a constrained
problem and leads us to the following states: \eqn{macrostates}{ |
j, \theta ,\varphi\rangle = \sum_{k=1}^{2j+1}
2^{-j} \sqrt{\left( \begin{array}{c} 2j\\
k-1 \end{array} \right)} (1+\cos \theta )^{\frac{j-k+1}{2}}
(1-\cos \theta )^{\frac{k-1}{2}} e^{\imath (k-1) \varphi}  \mid j,
j-k+1 \rangle } These obey
$\<j,\theta,\phi|j,\theta,\phi\>=1$ and (\ref{polar}) with
\eqn{rj}{ r=\sqrt{\<x\>\cdot\<x\>}=\lambda j, \quad \delta =
\frac{1}{j}.} We see that in these states the `true radius'
$|\<x\>|$ is $\lambda j$. The square root of the Casimir does not
give this true radius since it contains also the uncertainty
expressed in the variance of the position operators, but the error
$\delta$ vanishes as $j\rightarrow \infty$. Thus the larger the
representation, the more the geometry resembles to the classical.

We can therefore use these states $|j,\theta,\phi\>$ to convert
noncommutative geometric functions $f(x)$ into classical ones  in
spherical polar coordinates defined by \eqn{vevf}{
\<f\>(r,\theta,\phi)\equiv\<j,\theta,\phi|f(x)|j,\theta,\phi\> }
where $r=\lambda j$ is the effective radius. If we start with a
classical function $f$ and insert noncommutative variables in some
order then $\<f(x)\>$ (which depends on the ordering) looks more
and more like $f(\<x\>)$ as $j\to \infty$ and $\lambda\to 0$ with
the product fixed to an arbitrary $r$. 
As an example, the noncommutative
spherical harmonics $Y_l{}^m$ in Section~3 are already ordered in
such a way that replacing the noncommutative variables by the
expectation values $\<x_a\>$ gives something proportional to the
classical spherical harmonics. On the other hand $\<Y_l{}^m\>$
vanish  for $l> 2j$ and only approximate the classical ones for
lower $l$. Moreover, in view of the above, we expect \eqn{vevdf}{
\<\del_i f\>=\bar\del_i \<f\>+O(\lambda, {1\over j})} where
$r=j\lambda$ and  $\bar\del_i$ are the classical derivatives in
the polar form \bray {\bar{\partial}}_1 &=& \sin \theta \cos
\varphi \frac{\partial}{ \partial r} + \frac{1}{r} \cos \theta
\cos \varphi \frac{\partial}{\partial \theta} - \frac{1}{r } \sin
\theta\, \sin \varphi
\frac{\partial}{\partial \varphi} , \nonumber\\
{\bar{\partial}}_2 &=& \sin \theta \sin \varphi
\frac{\partial}{\partial r} + \frac{1}{r} \cos \theta \sin \varphi
\frac{\partial}{\partial \theta} + \frac{1}{ r } \sin \theta\,
\cos \varphi
\frac{\partial}{\partial \varphi} , \nonumber\\
{\bar{\partial}}_3 &=& \cos \theta \frac{\partial}{\partial r}
-\frac{1}{r} \sin \theta \frac{\partial}{\partial \theta} .
\nonumber \eray where we understand ${\del\over\del r}={1\over
\lambda}{\del\over\del j}$ on expectation values computed as
functions of $j$. More precisely one should speak in terms of the
joint limit as explained above with $\lambda j=r$ a continuous
variable in the limit. We note finally that the recent
star product for $\RL$ pointed out to us in \cite{HLS} suggests that
it should be possible to extend such a semiclassical 
analysis to all orders.

\appendix
\section{2-D and 3-D Calculi on $\RL$}

It might be asked why we need to take a four dimensional calculus
on $\RL$ and not a smaller one. In fact bicovariant differential
calculi on enveloping algebras $U({\mathfrak g})$ such as
$\RL\isom\U$ have been essentially classified\cite{majid2} and in
this appendix we look at some of the other possibilities for our
model. In general the coirreducible calculi (i.e. having no proper
quotients) are labelled by pairs $(V_{\rho} , \Lambda )$, with
$\rho : U ({\mathfrak g}) \rightarrow \mbox{End} V_{\rho}$ an
irreducible representation of $U ({\mathfrak g})$ and $\Lambda$ a
ray in $V_{\rho}$. In order to construct an ideal in
$\ker\epsilon$, take the map \[ \rho_{\Lambda} : U ({\mathfrak g})
\rightarrow  V_{\rho},\quad  h \mapsto \rho (h)\cdot \Lambda .\]
It is easy to see that $\ker \rho_{\Lambda}$ is a left ideal in
$\ker\epsilon$. Then, if $\rho_{\Lambda}$ is surjective, the space
of 1-forms can be identified with $V_{\rho} =\ker \epsilon /\ker
\rho_{\Lambda}$. The general commutation relations are
\begin{equation}
\label{av}
a v =va + \rho (a)\cdot v ,
\end{equation}
and the derivative for a general monomial $\xi_1 \ldots \xi_n$ is
given by the expression
\[
\extd(\xi_1 \ldots \xi_n ) = \sum_{k=1}^{n} \sum_{\sigma \in
S_{(n,k)}} \rho_{\Lambda} ( \xi_{\sigma (1)} \ldots \xi_{\sigma
(k)} ) \xi_{\sigma (k+1)} \ldots \xi_{\sigma (n)} ,
\]
the sum being for all $(n,k)$ shuffles.

We explore some examples of coirreducible calculi for the
universal enveloping algebra $\RL$, generated by $x_{\pm}$ and $h$
satisfying the commutation relations (\ref{lambdacommutations2}).
First, let us analyse the three dimensional, coirreducible
calculus on $\RL$ by taking $V_{\rho} =\C^3$, with basis
\[
e_+ =\left( \begin{array}{c} 1\\ 0 \\ 0 \end{array} \right) ,
\quad e_0 =\left( \begin{array}{c} 0\\ 1 \\ 0 \end{array} \right)
, \quad e_- =\left( \begin{array}{c} 0\\ 0 \\ 1 \end{array}
\right) .
\]
In this basis, the representation $\rho$ takes the form
\[
\rho (x_+ ) = \l \left(
\begin{array}{ccc}
0 & 2 & 0 \\
0 & 0 & 1 \\
0 & 0 & 0
\end{array} \right) , \quad
\rho (x_- ) = \l \left(
\begin{array}{ccc}
0 & 0 & 0 \\
1 & 0 & 0 \\
0 & 2 & 0
\end{array} \right) , \quad
\rho (h) = \l \left(
\begin{array}{ccc}
2 & 0 & 0 \\
0 & 0 & 0 \\
0 & 0 & -2
\end{array} \right) .
\]
We choose, for example, $\Lambda =e_0$. The space of 1-forms
will be generated by the vectors $e_+$, $e_-$ and $e_0$.
The derivatives of the generators of the algebra are given by
\[
\extd x_+ =\l^{-1} \rho (x_+ ) \cdot e_0 = 2e_+ , \quad \extd x_-
=\l^{-1} \rho (x_- ) \cdot e_0 =2e_- , \quad \extd h =\l^{-1} \rho
(h) \cdot e_0 =0 .
\]
The commutation relations between the basic 1-forms and the
generators can be deduced from (\ref{av}) giving
\bray
\label{redrelations1}
x_+ e_+ &=& e_+ x_+ ,\nonumber\\
x_+ e_0 &=& e_0 x_+ + 2\l e_+ , \nonumber\\
x_+ e_- &=& e_- x_+ +\l e_0  ,\nonumber\\
x_- e_+ &=& e_+ x_- +\l e_0 ,\nonumber\\
x_- e_0 &=& e_0 x_- + 2\l e_- , \nonumber\\
x_- e_- &=& e_- x_-  ,\nonumber\\
h e_+ &=& e_+ h + 2\l e_+ ,\nonumber\\
h e_0 &=& e_0 h ,\nonumber\\
h e_- &=& e_- h - 2\l e_- . \eray The expression for the
derivative of a general monomial $x_+^a x_-^b h`^c$ is \bray
\label{deriv1} \extd(x_+^a x_-^b h^c ) &=& 2ae_+  x_+^{a-1} x_-^b
h^c
+2be_-  x_+^a x_-^{b-1} h^c +\nonumber\\
&+&  2\l abe_0 x_+^{a-1} x_-^{b-1} h^c + 4\l^2 a(a-1)be_+
x_+^{a-2} x_-^{b-1} h^c . \eray We define the exterior algebra by
skew-symmetrizing and using similar methods as in Section~4 we
compute the  cohomologies as:
\[
H^{0} =\C  [h], \quad H^{1} = e_0 \C  [h], \quad H^{2}
=H^{3} =\{ 0 \}.
\]
This calculus is a three dimensional calculus but we have
introduced an isotropy by choosing $\Lambda$, and related to this
all functions of $h$ are killed by $\extd$, which is why the
cohomology is large.  This is why we do not take this calculus
even though it has the `obvious' dimension. There is the same
problem if we choose any other direction $\Lambda$.

We can also have a two dimensional coirreducible calculus on $\U$
using then $V_{\rho} =\C^2$, with basis
\[
e_1 =\left( \begin{array}{c} 1\\ 0  \end{array} \right) , \quad
e_2 =\left( \begin{array}{c} 0\\ 1  \end{array} \right) .
\]
In this basis, the representation $\rho$ takes the form
\[
\rho (x_+ ) =\l \left(
\begin{array}{cc}
0 & 1 \\
0 & 0
\end{array} \right) , \quad
\rho (x_- ) = \l \left(
\begin{array}{cc}
0 & 0 \\
1 & 0
\end{array} \right) , \quad
\rho (h) = \l \left(
\begin{array}{cc}
1 & 0  \\
0 & -1
\end{array} \right) .
\]
Choosing $\Lambda =e_1$, the space of 1-forms will be generated by
$e_1$ and $e_2$ and the derivatives of the generators of the
algebra are given by
\[
\extd x_+ =\l^{-1} \rho (x_+ ) \cdot e_1 =0 , \quad \extd x_-
=\l^{-1} \rho (x_- ) \cdot e_1 =e_2 , \quad \extd h =\l^{-1} \rho
(h) \cdot e_1 =e_1 .
\]
The commutation relations between the basic 1-forms and the
generators are then
\bray
\label{redrelations2}
x_+ e_1 &=& e_1 x_+ ,\nonumber\\
x_+ e_2 &=& e_2 x_+ + \l e_1 , \nonumber\\
x_- e_1 &=& e_1 x_- +\l e_2 ,\nonumber\\
x_- e_2 &=& e_2 x_-  ,\nonumber\\
h e_1 &=& e_1 h + \l e_1 ,\nonumber\\
h e_2 &=& e_2 h - \l e_2 . \eray And the derivative of a monomial
$x_-^a h^b x_+^c$ is given by \eqn{deriv2}{ \extd(x_-^a h^b x_+^c
) = e_1 \left( \sum_{i=0}^{b} \left(
\begin{array}{c}
b\\
i
\end{array}
\right) \l ^{i-1} x_-^a h^{b-i} x_+^c  \right) + e_2 \left(
\sum_{i=0}^{b} (-1)^i \left(
\begin{array}{c}
b\\
i
\end{array}
\right) \l^i a
 x_-^{a-1}h^{b-i} x_+^{c}  \right).} The cohomology of this calculus comes out as
\[
H^{0} =\C  [x_+ ],\quad  H^{1} =H^{2} =\{ 0 \} .
\]
Here again $\extd$ vanishes on all functions of $x_+$, which is related to
our choice of $\Lambda$. On the other hand this calculus motivates
us similarly to take for $\rho$ the tensor product of the spin $1\over
2$ representations and its dual. In this tensor product
representation there is a canonical choice of $\Lambda$, namely
the $2\times 2$ identity matrix. This solves the anisotropy and
kernel problems and this is the calculus that we have used on $\RL$ as the
natural choice in our situation. The above spinorial ones are 
coirreducible quotients of it.

\end{document}